\documentclass[aps,prl,twocolumn,longbibliography,showpacs,amsmath,amsmath,amssymb,amsfonts,superscriptaddress]{revtex4-1}
\usepackage{graphicx}
\usepackage{dcolumn}
\usepackage{bm}
\usepackage{color}
\usepackage{multirow}
\usepackage{hyperref}
\usepackage[normalem]{ulem}

\newcommand{\lan}{\left\langle}
\newcommand{\ran}{\right\rangle}

\newcommand{\mal}{\mathcal}

\newcommand{\ha}{\frac{1}{2}}

\newcommand{\ket}[1]{\left| #1 \ran}
\newcommand{\bra}[1]{\lan #1 \right|}

\newcommand{\comm}[2]{\left[ #1, #2 \right]}
\newcommand{\acomm}[2]{\left\{ #1, #2 \right\}}

\begin{document}
\title{Quench dynamics of a dissipative Rydberg gas in the classical and quantum regimes}

\author{Dominic Gribben}
\affiliation{School of Physics and Astronomy, University of Nottingham, Nottingham, NG7 2RD, United Kingdom}
\affiliation{Centre for the Mathematics and Theoretical Physics of Quantum Non-equilibrium Systems,
University of Nottingham, Nottingham NG7 2RD, UK}
\author{Igor Lesanovsky}
\affiliation{School of Physics and Astronomy, University of Nottingham, Nottingham, NG7 2RD, United Kingdom}
\affiliation{Centre for the Mathematics and Theoretical Physics of Quantum Non-equilibrium Systems,
University of Nottingham, Nottingham NG7 2RD, UK}
\author{Ricardo Guti\'errez}
\affiliation{School of Physics and Astronomy, University of Nottingham, Nottingham, NG7 2RD, United Kingdom}
\affiliation{Centre for the Mathematics and Theoretical Physics of Quantum Non-equilibrium Systems,
University of Nottingham, Nottingham NG7 2RD, UK}
\affiliation{Complex Systems Group \& GISC, Universidad Rey Juan Carlos, 28933 M\'{o}stoles, Madrid, Spain}

\keywords{}
\begin{abstract}
Understanding the non-equilibrium behavior of quantum systems is a major goal of contemporary physics. Much research is currently focused on the dynamics of many-body systems in low-dimensional lattices following a quench, i.e., a sudden change of parameters. Already such a simple setting poses substantial theoretical challenges for the investigation of the real-time post-quench quantum dynamics. In classical many-body systems the Kolmogorov-Mehl-Johnson-Avrami model describes the phase transformation kinetics of a system that is quenched across a first order phase transition. Here we show that a similar approach can be applied for shedding light on the quench dynamics of an interacting gas of Rydberg atoms, which has become an important experimental platform for the investigation of quantum non-equilibrium effects. We are able to gain an analytic understanding of the time-evolution following a sudden quench from an initial state devoid of Rydberg atoms and identify strikingly different behaviors of the excitation growth in the classical and quantum regimes. Our approach allows us to describe quenches near a non-equilibrium phase transition and provides an approximate analytic solution deep in the quantum domain.
\end{abstract}


\maketitle

When a liquid is cooled below the melting point, crystalline nuclei will appear, grow and eventually span the whole system. The volume fraction that has transformed into the crystalline solid at a given time, $f(t)$, is the natural macroscopic observable for the study of the kinetics of first order phase transitions. The standard stochastic model of nucleation and growth, the Kolmogorov-Johnson-Mehl-Avrami (KJMA) theory  \cite{kolmogorov1937,avrami1939,johnson1939,avrami1940,avrami1941}, is of widespread use in metallurgy and material science \cite{christian2002}. This model predicts a compressed exponential form $f(t) = 1 - \exp{\left[-(t/\tau)^n\right]}$, the so-called Avrami equation, where  $n=d+1$ in a $d$-dimensional system with constant (homogeneous) nucleation rate and ballistic growth. The processes considered in this formulation are the nucleation of solid domains, their growth, and the coalescence of expanding domains \cite{jun2005a}. Recently, the KJMA model has found applications in the study of DNA replication \cite{jun2005b}, epidemic spreading in networks \cite{avramov2007}, and the melting of stable glasses \cite{gutierrez2016a,jack2016}, to name but a few examples.

In this work, we show that the KJMA picture of phase transformation kinetics allows to quantitatively understand the non-equilibrium dynamics of an open many-body quantum system that is subjected to a sudden quench. While most studies on quenches  focus on closed systems, see e.g. the theoretical work in Refs. \cite{rigol2006, cazalilla2006,calabrese2006,kollath2007,lauchli2008,mitra2017} as well as experiments realized with ultracold atomic gases or trapped ions \cite{kinoshita2006,hofferberth2007,chen2011,gring2012, schreiber2015,kaufman2016,schachenmayer2013,richerme2014}, the considered scenario complements a growing number of recent contributions to the study quenches in dissipative dynamics \cite{kashuba2013,kennes2013,creatore2014,henriet2016,shapourian2016,bernier2017}. A natural platform for exploring open system quenches are gases of highly excited Rydberg atoms \cite{gallagher2005,low2012}, which allow to tune the relative strength of coherent and classical processes and thereby the degree to which the many-body system is open. Atoms in Rydberg states are strongly interacting and the interplay between this interaction and their laser excitation is the source of collective effects. Of particular current interest is so-called facilitated excitation, where the laser frequency is chosen such that the excitation of an atom in the vicinity of an already excited one is enhanced \cite{amthor2010,lesanovsky2014, marcuzzi2017,simonelli2016,urvoy2015,schempp2014}. Upon a sudden quench from an initial state devoid of Rydberg excitations, mechanisms analogous to the two basic processes of the nucleation-and-growth KJMA model govern the subsequent relaxation dynamics: isolated spontaneous excitations (seeds) act as nuclei, from which transformed domains grow due to facilitation. In the effective classical limit --- where the Rydberg excitation is an incoherent process \cite{ates2007,lesanovsky2013,garttner2013,marcuzzi2014,sibalic2016} --- the evolution of the density of excitations is captured by an Avrami equation with diffusive growth. In the opposite quantum coherent limit relaxation proceeds through coherently evolving pairs of domain walls that propagate ballistically. Remarkably, the Avrami approach enables an approximate analytic solution of the non-equilibrium many-body dynamics in this quantum regime.

\begin{figure*}[t]
\includegraphics[scale=0.55]{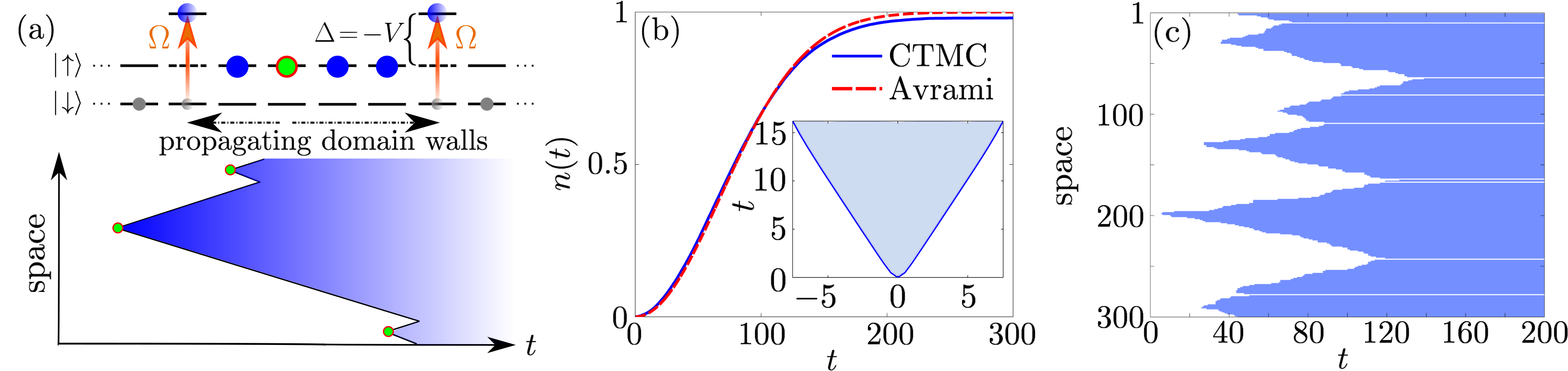}
\caption{ {\sf \bf Illustration of idealized quench dynamics.}
(a) Domains of excited atoms (blue) grow from spontaneous seeds (green). In a Rydberg gas this is achieved via facilitated excitation (see text) that is  driven by a laser with Rabi frequency $\Omega$. (b) Excitation density as a function of time $n(t)$ (blue) for a system of $N=100000$ (average based on 10 realizations) and $R=2$, where the de-excitation process is artificially switched off, and analytical prediction given by the Avrami equation in Eq. \ref{avraminodeex} (red). The inset shows the dynamics of the domain boundaries (see blue line) starting from a configuration with a single excitations in the center of the chain (based on 10000 realizations). The blue color of the region contained within the boundaries highlights the transformed domain.  (c) Representative trajectory of a chain of $N=300$ atoms for $R=2$.}\label{fig1}
\end{figure*}

\noindent {\bf \em Elementary processes and quench protocol.} We consider a lattice gas where each site contains a single atom, and focus on 1D chains. In this setting a deeper understanding of the KJMA model exists \cite{sekimoto1984,sekimoto1986,sekimoto1991,benami1996,jun2005a}, and the quench dynamics can be most conveniently explored in optical lattice quantum simulators \cite{bloch2012,labuhn2016,bernien2017,lienhard2017,zeiher2017}. Atoms can either be in their ground state $\ket{\downarrow}$ or in a high-lying (Rydberg) excited state $\ket{\uparrow}$. The interaction $V_{kl}$  between excited atoms $k$ and $l$ depends on the distance, typically with a power law form $|r_k - r_l|^{-\alpha}$. The transition between the two atomic states is driven by a laser field with Rabi frequency $\Omega$, and detuning $\Delta$ w.r.t. the atomic resonance frequency. The coherent part of the dynamics is thus generated by the Hamiltonian
\begin{equation}
H = \Omega \sum_k \sigma_k^x +\Delta \sum_k n_k+ \sum_{k<l} V_{kl} n_k n_l,
\label{hamiltonian}
\end{equation}
where $n_k = \ket{\uparrow}_k \bra{\uparrow}$ and $\sigma_k^x = \sigma_k^+ + \sigma_k^-$ for $\sigma_k^+ = \ket{\uparrow}_k \bra{\downarrow}$ and $\sigma_k^- = \ket{\downarrow}_k \bra{\uparrow}$. Throughout we consider that the system is excited under facilitation conditions, i.e. the excitation process is resonant next to an already excited atom. This is achieved by setting the detuning such that it cancels the interaction energy of adjacent excited atoms [see Fig. \ref{fig1} (a), upper panel]: $\Delta =  - V_{k,k+1} \equiv- V$ \cite{amthor2010,lesanovsky2014,helmrich2016,marcuzzi2017,letscher2017}. It is important to note, that there is still a small probability for unfacilitated (spontaneous) excitation. Dissipative processes we consider are dephasing (through laser linewidth, thermal effects, etc. \cite{low2012}) with rate $\gamma$ and spontaneous radiative decay of excited atoms, which occurs with a rate $\kappa$. These effects are accounted for by dissipators $\mal{L}(J) \rho = J \rho J^\dag - \ha \acomm{J^\dag J}{\rho}$ with jump operators $J^{\text{(deph)}} = \sqrt{\gamma}\, n_k$ and $J^{\text{(dec)}} = \sqrt{\kappa}\, \sigma_k^-$. Including both incoherent and coherent processes, the evolution of the density matrix $\rho$ is governed by the Lindblad equation
\begin{equation}
	\partial_t \rho = - i\comm{H}{\rho} + \sum_k \left( \mal{L}(\sqrt{\gamma}\, n_k) + \mal{L}(\sqrt{\kappa}\, \sigma_k^-)   \right) \rho.
\label{qmastereq}
\end{equation}

The quench protocol is as follows. We take as initial state the stationary state for $\Omega=0$ (laser off), which is the `empty' configuration $\ket{\downarrow \downarrow \cdots \downarrow}$. We then change to $\Omega > 0$ (laser on), and let the system evolve towards the new stationary state, which in general will contain a finite density of excitations. In an idealized description of the dynamics, the evolution following the quench is characterized by four basic processes:  (i) slow unfacilitated excitation $\ket{\downarrow \downarrow} \to \ket{\downarrow \uparrow}$, which creates new excitation domains (ii) fast facilitated excitation $\ket{\uparrow \downarrow} \to \ket{\uparrow \uparrow}$, which can lead to domain growth by the propagation of excitations (atom $k$ facilitates atoms $k\pm 1$, which in their turn facilitate $k\pm 2$, etc.), (iii) decay $\ket{\uparrow} \to \ket{\downarrow}$, which introduces an imbalance between excitation and de-excitation in favor of the latter, and (iv) dephasing $(\ket{\uparrow} + \ket{\downarrow})(\bra{\uparrow} + \bra{\downarrow}) \to \ket{\uparrow}\bra{\uparrow} + \ket{\downarrow}\bra{\downarrow}$, which determines the nature of the dynamics, ranging from quantum coherent to classical stochastic. Processes (i) and (ii) are analogous to nucleation and growth in the KJMA framework \cite{kolmogorov1937,avrami1939,johnson1939,avrami1940,avrami1941,christian2002}, respectively. This list does not exhaust all dynamical possibilities,  but allows us to highlight dominant processes that will be referred to frequently in the following sections. In practice, there will be other transitions as well, e.g. unfacilitated and facilitated de-excitations, which occur with the same rates as the reverse processes. They will introduce some new features that are absent in the classical KJMA picture, as will be discussed below.

\noindent{\bf \em Classical limit: ballistic and diffusive behavior.} Current experiments frequently operate in the limit of strong dephasing (see, e.g., \cite{schempp2014,urvoy2015,valado2016}). Here, the dynamics is governed by a (classical) stochastic rate equation, where transitions $\ket{\downarrow}_k\leftrightarrow\ket{\uparrow}_k$ have associated configuration-dependent rates
\begin{equation}
\Gamma_k=  \frac{4 \Omega^2}{\gamma}\frac{1}{1+R^{2 \alpha}\left(1 -   \sum_{l\neq k} \frac{n_l}{|r_l - r_k|^\alpha} \right)^2},
\label{rate}
\end{equation}
with interaction parameter $R^\alpha = 2 V/\gamma$ \cite{lesanovsky2013,marcuzzi2014}. The functional form of the rates implies that unfacilitated excitations occur with rate $\Gamma_\text{spon} = \frac{4\Omega^2}{\gamma}/\left(1+R^{2\alpha}\right)$, and facilitated excitations, with rate $\Gamma_\text{fac} = \frac{4\Omega^2}{\gamma}$. The same rates apply to unfacilitated and facilitated de-excitations, respectively, which will be initially left out of the dynamics in the numerical explorations below, in order to highlight the connection between the physics under study and the KJMA theory. In the following, we consider van der Waals interactions, $\alpha = 6$, and rescale time by the facilitation timescale $4 \Omega^2/\gamma$.

After the quench (i.e. after switching on the excitation laser, such that $\Omega > 0$), spontaneous excitations start to appear with rate $\Gamma_\text{spon}$. From these seeds, facilitating processes originate that excite neighboring sites, thus starting an excitation front that leads to the growth of transformed domains. When the interaction parameter $R$ is sufficiently large, the spontaneous excitation rate is much smaller than that rate at which a domain grows (i.e. $\Gamma_\text{spon}/\Gamma_\text{fac}\ll 1$), see Fig. \ref{fig1} (a) upper panel, making the evolution reminiscent of the nucleation and growth problem that is described by the KJMA theory. We thus expect to be able to capture the macroscopic evolution with a transformed fraction $f(t)$, defined as the fraction of the atoms that have undergone at least one excitation process after the quench. Following the KJMA arguments
\cite{kolmogorov1937,avrami1939,johnson1939,avrami1940,avrami1941}, one  should expect $f(t) = 1 - e^{-2 \Gamma_\text{spon} \int_0^t d\tau\, G(t-\tau)}$, where $\Gamma_\text{spon}$ acts as nucleation (seed creation) rate, and $G(t)$ determines the growth law of a domain. A derivation can be found in the Supplemental Material \cite{SM}.

\begin{figure}[t]
\hspace{-0.3cm}\includegraphics[scale=0.42]{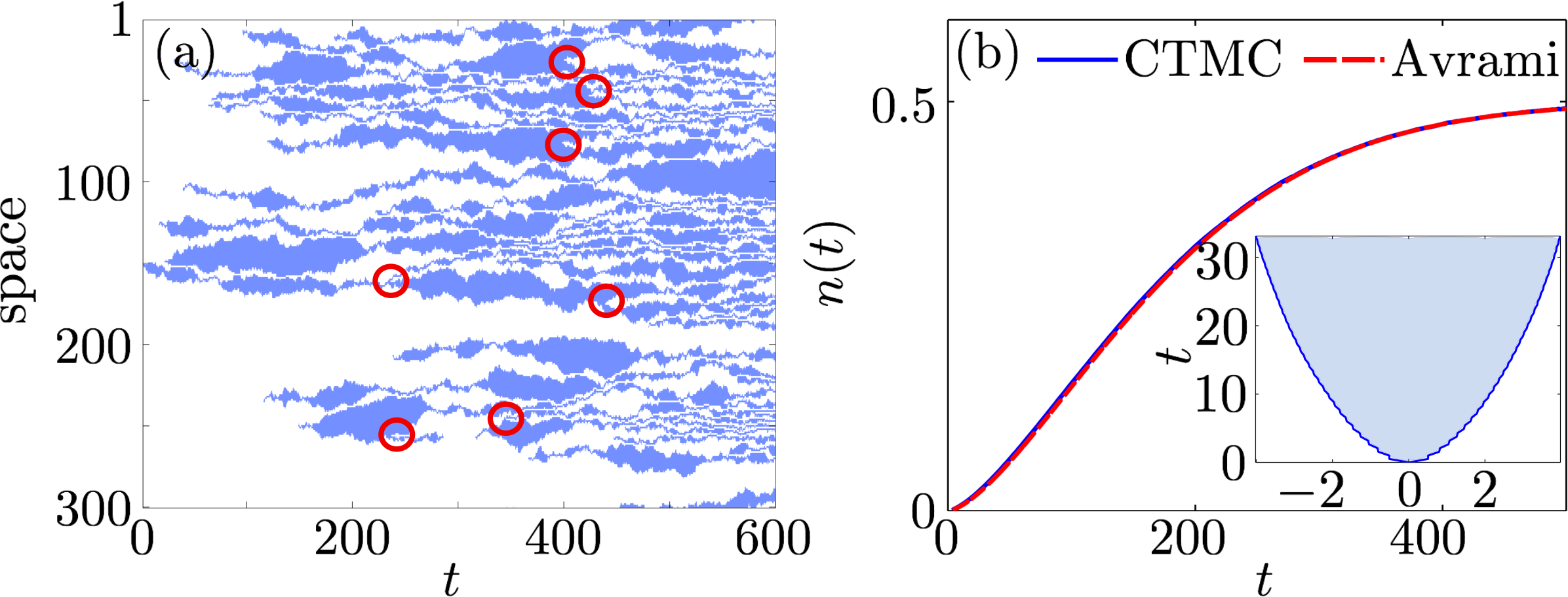}
\caption{ {\sf \bf Diffusive regime.} (a) Representative trajectory of a chain of $N=300$ atoms. Red circles highlight ``spontaneous de-excitations'' within domains, from which de-excitation fronts originate (see text for an explanation).  (b) Numerically calculated excitation density $n(t)$ (blue) for a system with $N=1000$ atoms (average based on 10000 realizations) and analytic Avrami curve (red). The inset shows the evolution of the boundaries of a single domain emerging from a seed at the center of the chain (see blue line, 10000 realizations have been averaged). The blue color of the region contained within the boundaries highlights the transformed domain.}\label{fig2}
\end{figure}

In a first step, we consider an idealized dynamics without de-excitations and without decay (i.e. only $\ket{\downarrow}\to\ket{\uparrow}$ transitions are possible), which is relevant for the very initial stages after the quench \cite{gutierrez2015}. In this case, the front arising from an unfacilitated excitation, $\ket{\downarrow \downarrow \downarrow \uparrow \downarrow \downarrow \downarrow} \to \ket{\downarrow\downarrow \uparrow \uparrow \uparrow \downarrow\downarrow} \to \ket{\downarrow\uparrow \uparrow \uparrow \uparrow \uparrow\downarrow}$, is expected to propagate ballistically, see Fig. \ref{fig1} (a). This is confirmed by continuous-time Monte Carlo (CTMC) simulations \cite{bortz1975} starting from a single initial excitation, see the inset of Fig. \ref{fig1} (b).  In Fig. \ref{fig1} (c), we see a typical trajectory, where several nucleation events are observed to give rise to facilitation fronts. Due to the power-law interactions in Eq. (\ref{rate}), the time it takes to facilitate an excitation at the boundary of a large domain, $\ket{\cdots \uparrow \uparrow \uparrow \uparrow \downarrow \downarrow \downarrow \downarrow} \to \ket{\cdots\uparrow \uparrow \uparrow \uparrow \uparrow \downarrow\downarrow \downarrow}$, is given by $\tilde{\Gamma}^{-1}\approx 1 + \mathcal{S}_2^2\, R^{12}$, where $\mathcal{S}_2 \equiv \sum_{l=2}^\infty l^{-6} =\frac{\pi^6}{945}-1\approx 0.017$. We thus have the domain growth law $G(t) = \tilde{\Gamma} t$, which leads to
\begin{equation}
f(t) = 1 - e^{-\Gamma_\text{spon}  \tilde{\Gamma}\, t^2}.
\label{avraminodeex}
\end{equation}
The transformed fraction is equal to the density of excitations, $f(t) = n(t) \equiv N^{-1} \sum_k \langle n_k(t) \rangle$, as domains contain only up-spins in this case, but the distinction will be important below when we consider de-excitations. In Fig. \ref{fig1} (b) we compare Eq.\! (\ref{avraminodeex}) with the density of excitations  in the CTMC simulations. The agreement is excellent showing the applicability of the KJMA approach. The small discrepancy at long times results from the fact that the boundaries of merging domains excite only at very long times, as their associated rates are approximately $\Gamma_\text{spon} \ll \tilde{\Gamma}$ [see Fig. \ref{fig1} (c)].

In the presence of de-excitation processes the domains emerging from seeds acquire a more complex shape, see Fig. \ref{fig2} (a). Domains grow now diffusively, as each boundary of a large domain behaves as a random walker. The probability to move one site away from or towards the nucleation center is approximately the same. This is clearly observed by initializing the system with a single excitation, see the inset of Fig. \ref{fig2} (b). The diffusion timescale is $\bar{\Gamma}^{-1} = 1 + b^2 \mathcal{S}^2_2 R^{12}$, where $b\in [1/2,  1]$ is the (unknown) domain density in the vicinity of the boundaries. The probability of a boundary to be at position $k$ evolves according to $\partial_t P(k,t) = \bar{\Gamma} \left[P(k-1,t)+P(k+1,t)-2P(k,t)\right]$, which in the continuum limit gives $\partial_t P(x,t) = \bar{\Gamma}\, \partial_x^2 P(x,t)$. Then $\partial_t \langle x^2 \rangle = 2\, \bar{\Gamma}$, and we obtain for the domain growth function $G(t) = \sqrt{\langle x^2 \rangle} = \sqrt{2\, \bar{\Gamma}\, t}$. This yields the Avrami function
\begin{equation}
f(t) = 1 - e^{-\frac{4 \sqrt{2}}{3} \Gamma_\text{spon} \bar{\Gamma}^{1/2}\, t^{3/2}},
\label{avramifull}
\end{equation}
where the power of the time dependence has changed w.r.t. Eq. (\ref{avraminodeex}). This expression gives us the fraction of the system that has been reached by the excitation domain boundaries. However, by inspecting Fig. \ref{fig2} (a) one can see that formed domains do not remain unaltered when time progresses. Rather, ``spontaneous de-excitations'' [see red circles in Fig. \ref{fig2} (a)] occur within them that lead to de-excitation fronts, giving rise to secondary (de-excitation) nucleation and growth processes within the domains. In fact, as both the spontanous and the facilitated de-excitation rates in a fully excited domain are very similar to the excitation rates previously considered [see Eq. (\ref{rate})], such de-excitation processes must evolve in time in much the same way as the excitation processes that are the main object of our study. While a full characterization of the combined effect of such secondary and higher order (excitation and de-excitation) processes appears to be challenging, on average their effect is captured by simply assuming that domains instantaneously reach the stationary density of $1/2$, $n(t) \approx f(t)/2$ --domains lose excitations on the same timescales along which new domains form. In Fig. \ref{fig2} (b), we plot $n(t)$ obtained via CTMC and $f(t)/2$ as in Eq. (\ref{avramifull}) choosing the free parameter $b=0.68$ [using Eq. (\ref{avramifull}) to fit the numerically obtained $n(t)$ with $b$ as fitting parameter yields $b = 0.68 \pm 0.05$], which show an excellent agreement.

\begin{figure}[t!]
\hspace{-0.2cm}\includegraphics[scale=0.37]{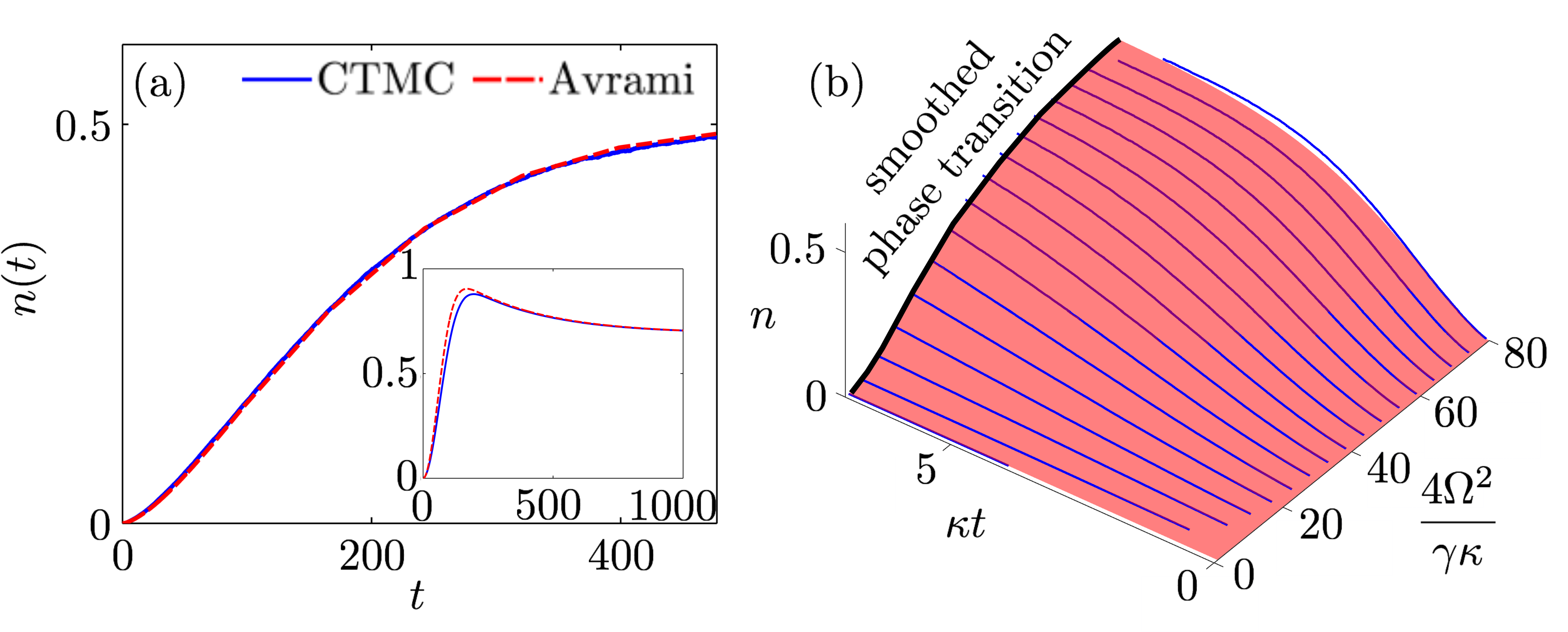}
\caption{ {\sf \bf Quench dynamics in the presence of decay.}
(a) Density of excitations $n(t)$ (blue) for a system of $N=1000$ (average based on 1000 realizations), and Avrami curve (red), for decay rate $\kappa = 10^{-3}$. Inset: analogous results for the dynamics without de-excitations (axis labels are the same as in the main panel). (b) Excitation density $n$ as a function of normalized time $\kappa t$ and relative driving strength $4\Omega^2/\gamma\kappa$: continuous time Monte Carlo results (blue lines), for $N=1000$, and Avrami curve (red surface). For long times the density approaches a 
(quasi-)stationary state (thick black line) which displays a 
smoothed non-equilibrium phase transition \cite{marcuzzi2015,gutierrez2016c} as a function of the decay rate. Averages based on 1000 realizations.}\label{fig3}
\end{figure}

\noindent{\bf \em Classical limit: quench dynamics in the presence of decay.}  For very long times, radiative decay of excited atoms to their ground state becomes an important element of the dynamics. For small decay rate, i.e. when $\kappa \ll \bar{\Gamma},\tilde{\Gamma}$, the domain growth law $G(t)$ will not be modified significantly, but the density of excitations within each domain will decrease. This requires a modification of the KJMA approach: to this end we introduce a function $c(t)$ that represents the concentration of excitations in the transformed domains as a function of time, which will be estimated from CTMC simulations as explained below. The total excitation density is then given by the convolution of the transformation rate and the concentration, $n(t) = \int_0^t d\tau \dot{f}(\tau)\, c(t-\tau)$, where $c(t-\tau)$ accounts for the concentration at time $t$ of regions that were reached by the front at a time $\tau < t$. For $\kappa = 0$, we have $c(t) = 1/2$ for intermediate and long times ($c(t) = 1$ if de-excitations are suppressed), thus recovering the results of the previous section. Fig. \ref{fig3} (a) shows that this modelling excellently describes the data obtained from CTMC simulations. Here, $n(t)$ is based on the convolution of Eq. (\ref{avramifull}) and the concentration $c(t)$ resulting from the CTMC evolution of system starting from a random configuration with density $1/2$, which show an excellent agreement. We report analogous results for the dynamics without de-excitations in the inset.

More importantly, this approach allows for the exploration of the excitation density $n(t)$ as a function of the decay rate $\kappa$. The corresponding data are shown in Fig. \ref{fig3} (b). At stationarity (black line) $n(t)$ acquires a sigmoidal shape \cite{marcuzzi2015}, that interpolates between two distinct stationary states, which are linked by a smoothed-out non-equilibrium phase transition \cite{marcuzzi2015,marcuzzi2016,gutierrez2016c}. For large decay rate $\kappa$ domain growth is suppressed and the stationary state density is close to zero. However, for sufficiently small values of $\kappa$ a stationary state with large excitation density is quickly approached. Both regimes are well captured by the KJMA approach demonstrating its applicability also in the vicinity of a phase transition.

\begin{figure}[t]
\includegraphics[scale=0.135]{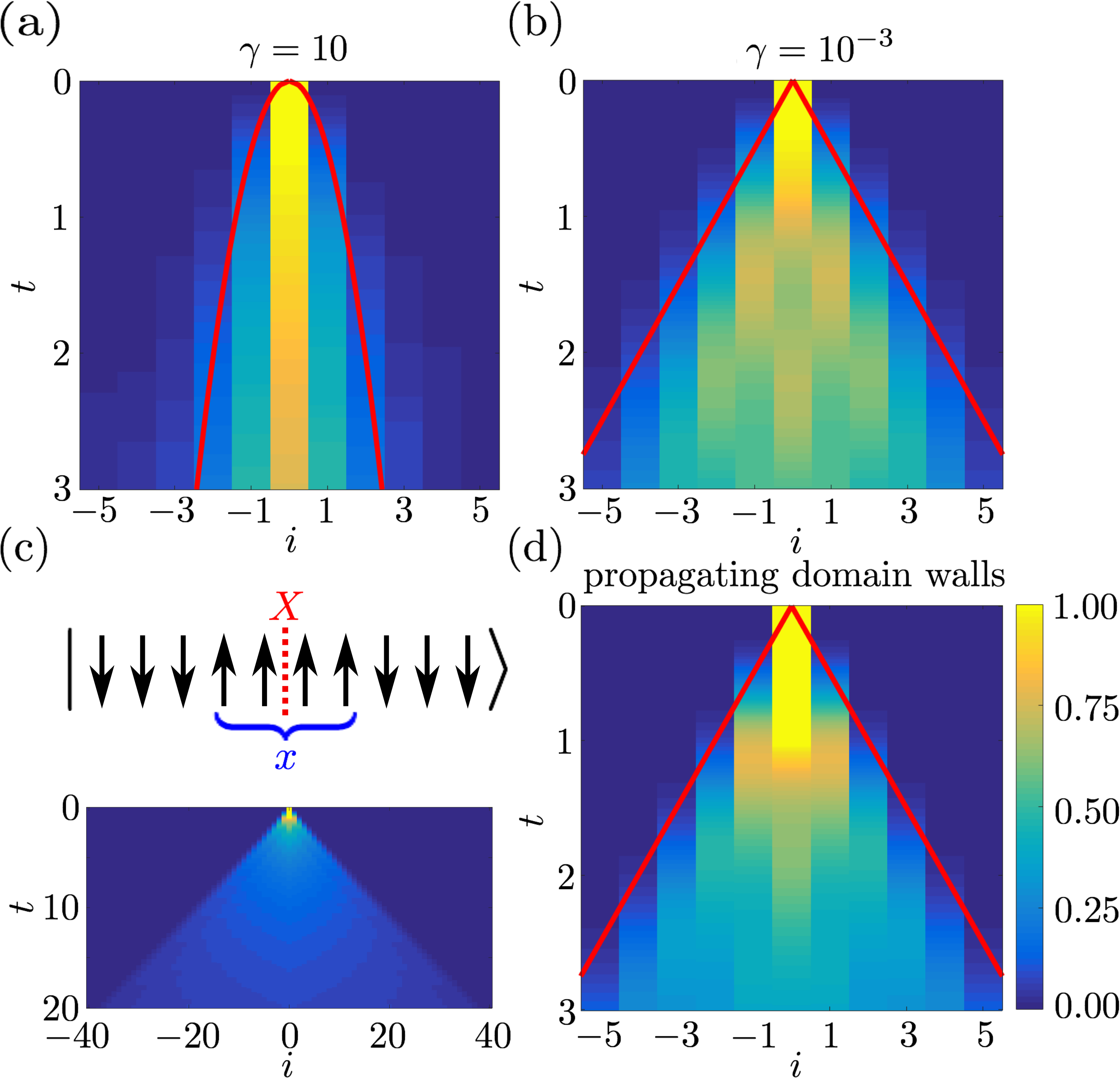}
\caption{ {\bf Domain growth in the quantum regime.}
(a) Local excitation density $\langle n_k(t)\rangle$ for $\gamma=10$ obtained via QJMC in a chain of $N=11$ atoms ($\Omega = 1$, $C_6 =-\Delta= 50$) and prediction based on the (classical) strongly dissipative dynamics (red line).
(b) Local excitation density $\langle n_k(t)\rangle$ for $\gamma=10^{-3}$ [other parameters as in (a)] obtained via QJMC and prediction based on the quantum domain wall propagation model, $G(t) = 2\Omega t$ (red line; see text). The results shown in (a) and (b) are averages of $1000$ QJMC runs. (c) Sketch illustrating the degrees of freedom  (upper plot) and $\langle n_k(t)\rangle$ as given by the solution of the domain wall model for a system of $N=81$ sites (lower plot). (d) $\langle n_k(t)\rangle$ as given by the solution of the domain wall model for a system of $N=11$ sites, and ballistic prediction [red line as in (b)].}
\label{Fig4}
\end{figure}

\noindent {\bf \em Quench dynamics in the quantum regime.}  In Rydberg experiments the quantum limit is approached by reducing the relative strength of the dephasing noise, such that $\gamma < \Omega$. In the following we show that the KJMA theory allows an approximate analytic solution for the quench dynamics in this regime. For simplicity, in this section we do not consider the effect of spontaneous decay. We begin by numerically simulating the evolution of a domain emerging from a single seed via quantum jump Monte Carlo (QJMC) simulations \cite{dalibard1992}, see Fig. \ref{Fig4} (a,b). For large dephasing, the domain walls propagate in a manner that seems compatible with our predictions of the strongly dissipative regime, i.e. the size of the domain appears to grow as $G(t)=\sqrt{2\bar{\Gamma} t}$, see red line in Fig. \ref{Fig4} (a).

In the quantum limit, we find a ballistic behavior, see Fig. \ref{Fig4} (b) for $\gamma = 10^{-3}\, \Omega$. In order to obtain an analytical expression of the growth function $G(t)$ we consider that an initial seed gives rise to two domain walls. Facilitation will only move the domain walls, but will neither create nor annihilate them. A domain is thus characterized by two coordinates: its length $x\geq 1$, and its center of mass $X$. The corresponding equation of motion is
\begin{equation}
i \partial_t \varphi_{x,X}\! =\!  \Omega\, (\varphi_{x+1,X+\frac{1}{2}} + \varphi_{x+1,X-\frac{1}{2}} + \varphi_{x-1,X+\frac{1}{2}} +\varphi_{x-1,X-\frac{1}{2}}).
\label{pairdomainwalls}
\end{equation}
As the domain grows or shrinks by one site, $X$ is shifted by $1/2$ to the left or to the right, depending on the affected boundary, see Fig. \ref{Fig4} (c), upper plot.  The solution of (\ref{pairdomainwalls}) can be explicitly written in terms of Bessel functions of the first kind, see  \cite{SM}. The maximum of the probability amplitude is reached at the position of the propagating wavefront, which follows $x\approx 4\Omega t + 1$. We thus obtain $G(t) = 2 \Omega t$, see red line in Fig. \ref{Fig4} (b) and (d). The time evolution corresponding to the solution of Eq. (\ref{pairdomainwalls}) (see \cite{SM} for details) is shown in  Fig. \ref{Fig4} (c) (lower panel) and (d).

To derive an expression for the transformed fraction $f(t)$, we need to combine the result for $G(t)$ with the rate for the spontaneous creation of excitations. The latter is a classical process, which takes place at a rate $2 \gamma \Omega^2/V^2$ (see Ref. \cite{SM}). This leads to
\begin{equation}
f(t) = 1 - e^{- \frac{4 \gamma \Omega^3}{V^2} t^2}, \label{q_avrami}
\end{equation}
which is clearly distinct from the Avrami curve in the classical regime. Note, that the KJMA approach giving rise to this expression is applicable only below a certain dephasing strength, as the ballistic growth takes place only for times smaller than $t_\text{deph} = 1/\gamma$. Eq. (\ref{q_avrami}) is thus valid only when the domain emerging from one seed hits the domains originating from its neighboring seeds [see Fig. \ref{fig1} (a)] at times smaller than $t_\text{deph}$. As the average density of spontaneously created seeds after time $t$ is $(2 \gamma \Omega^2/V^2)\, t$, the typical distance between seeds is the inverse of this quantity. On the other hand, the radii of emerging domains grow as $G(t) = 2\Omega t$, which must clearly exceed the typical distance between seeds when $t$ is approaching $t_\text{deph}$. Combining these constraints yields an upper bound for the dephasing rate $\gamma < 4\Omega^3/V^2$ below which Eq. (\ref{q_avrami}) is applicable.

\noindent {\bf \em Conclusions.} We have shown that KJMA framework serves as a basis for the analytical understanding of the quench dynamics of an open quantum system. While the quantum simulation of such systems is non-trivial even in 1D, modern Rydberg quantum simulators \cite{bernien2017,labuhn2016,bloch2012}  will be able to verify these predictions, in particular Eq. (\ref{q_avrami}). More importantly, experiments will be able to access very interesting regimes that are intermediate between 'classical' and 'quantum', and also will allow to probe quenches in higher spatial dimensions, for which numerically exact calculations become rapidly intractable. While we have focused on non-equilibrium Rydberg gases under facilitation conditions, our approach could be adapted to other cold atomic settings where relaxation is driven by nucleation events, as discussed e.g. in Ref. \cite{everest2017}.

\begin{acknowledgments}
We thank Federico Carollo for insightful discussions on the propagating domain wall model. The research leading to these results has received funding from the European Research Council under the European Union's Seventh Framework Programme (FP/2007-2013) / ERC Grant Agreement No. 335266 (ESCQUMA), the EPSRC Grant No. EP/M014266/1 and the H2020-FETPROACT-2014 Grant No. 640378 (RYSQ). IL gratefully acknowledges funding through the 
Royal Society Wolfson Research Merit Award. RG acknowledges the funding received from the European Union's Horizon 2020 research and innovation programme under the Marie Sklodowska-Curie grant agreement No. 703683. We are also grateful for access to the University of Nottingham High Performance Computing Facility.
\end{acknowledgments}


%

\onecolumngrid
\newpage

\renewcommand\thesection{S\arabic{section}}
\renewcommand\theequation{S\arabic{equation}}
\renewcommand\thefigure{S\arabic{figure}}
\setcounter{equation}{0}
\setcounter{figure}{0}

\begin{center}
{\Large SUPPLEMENTAL MATERIAL:\\\vspace{0.2cm} Quench dynamics of a dissipative Rydberg gas in the classical and quantum regimes}
\end{center}

\section{Avrami equation in one dimension}

While the Kolmogorov-Mehl-Johnson-Avrami model \cite{kolmogorov1937SM,avrami1939SM,johnson1939SM,avrami1940SM,avrami1941SM} has been studied in 1D, see e.g. Refs. \cite{sekimoto1984SM,sekimoto1986SM,sekimoto1991SM,benami1996SM,jun2005aSM}, here we rederive the corresponding Avrami equation giving the transformed fraction as a function of time $f(t)$ for completeness. This will also allow us to link the classical concepts of nucleation and growth to the excitation mechanisms existing in a dissipative Rydberg gas. We will assume the dynamics to take place in the real line, instead of a discrete lattice given by the positions of the atoms in the system, as the derivation is simpler in that setting. This will not compromise the validity of the Avrami equation, as microscopic details become immaterial in the analysis of macroscopic observables such as $f(t)$.

At time $t=0$, $n_\text{spon}(0)$ (homogeneously distributed) spontaneous excitations per unit length occur. The probability that the one that is the closest to the origin (or any other arbitrary point) lies at  a distance between $r$ and $r+dr$ is $p(r) dr = \left[1 - \int_0^r p(r^\prime) dr^\prime\right] 2 n_\text{spon}(0) \, dr$, where $1 - \int_0^r p(r^\prime) dr^\prime$ is the probability that the distance is not smaller than $r$, and the factor of $2$ arises from considering intervals of size $dr$ on both sides of the origin. By taking the derivative of $p(r)$, we obtain an ordinary differential equation that can be easily solved, yielding $p(r) dr = 2\, n_\text{spon}(0)\, e^{-2\, n_\text{spon}(0) r} dr$, where the prefactor $2\, n_\text{spon}(0)$ arises from the normalization condition $\int_0^\infty p(r) dr = 1$.
If the facilitation that starts from the spontaneous excitations is characterized by a domain growth function $G(t)$, the fraction of the system that remains untransformed by the nucleation and growth dynamics at time $t$ is $1 - f(t) = \int_{G(t)}^\infty p(r) dr = e^{-2 n_\text{spon}(0) G(t)}$. The domain growth function $G(t)$ represents the propagation of excitations via facilitation, and it is such that $G(t)=0$ for $t\leq0$, and $dG(t)/dt>0$ for $t>0$.

While facilitation leads to the expansion of previously created domains, further spontaneous excitations take place in the system,  giving rise to new domains. If, for simplicity, we consider a sequence of spontaneous excitations occurring at times $t_s = s \Delta t$ up to time $t$ (for $s=0, 1, \ldots, \mathcal{N}$, such that $\mathcal{N} = \lfloor t/\Delta t\rfloor$) with densities $n_\text{spon}(t_s)$ and associated growth functions $G(t-t_s)$, the untransformed fraction at time $t$ is the fraction of the system that has not been reached by any of the growing domains, $1 - f(t) = e^{-2 \sum_{s=0}^\mathcal{N} n_\text{spon}(s \Delta t) G(t-s \Delta t)}.$ In the limit of $\Delta t \to 0$, the transformed fraction becomes $f(t) = 1 - e^{-2 \int_0^t d\tau\, \dot{n}_\text{spon}(\tau) G(t-\tau)}$. Here, the nucleation rate $\dot{n}_\text{spon}(t)$ will always be constant, leading to the following Avrami equation
\begin{equation}
f(t) = 1 - e^{-2\, \dot{n}_\text{spon} \int_0^t d\tau\, G(t-\tau)}.
\label{avrami}
\end{equation}

In the classical regime, the nucleation rate is simply given by the spontaneous excitation rate, $ \dot{n}_\text{spon} = \Gamma_\text{spon}$ (see main text for definition), while the specific form of the growth function $G(t)$ will depend on the local processes included in the dynamics. The main steps underlying the derivation of the nucleation rate and the domain growth function in the quantum regime are provided in the remainder of this Supplemental Material.

\section{Coherent domain-wall model}

The transformation dynamics in the quantum regime can be understood by means of a domain-wall model, where the transformed domain can be characterized by two coordinates: the length of the domain $x\geq 1$, and its center of mass $X \in (-\infty,\infty)$. The unitary evolution of the system (in the absence of decoherence or spontaneous decay) is given by the Hamiltonian $H=\sum_k H_k$, where the term for site $k$ is given by $H_k = \Omega\, \sigma_k^x + (\sum_{l\neq k} V_{kl}\, n_l + \Delta)\, n_k$ for $\Delta =  - V_{k,k+1} \equiv- V$ (i.e. under facilitation conditions). For simplicity, we focus on the dominant nearest neighbour interactions, so the local Hamiltonian becomes $H_k = \Omega\, \sigma_k^x + V (n_{k+1} + n_{k-1} -1) n_k$, where the second term vanishes in the presence of an excited neighbor, and thus the hopping rate for a domain wall is given by $\Omega$. The evolution is therefore governed by 
\begin{equation}
i\, \partial_t \varphi_{x,X} =  \Omega \left[\varphi_{x+1,X+\frac{1}{2}} + \varphi_{x+1,X-\frac{1}{2}} + \varphi_{x-1,X+\frac{1}{2}} +\varphi_{x-1,X-\frac{1}{2}}\right].
\label{eqmot}
\end{equation}
As the domain grows or decreases by one site, the center of mass is shifted by $1/2$ to the left or to the right, depending on whether the change occurred on the left or the right boundary. If we make the coordinte change $y = 2 X$ and apply a shift, we can rewrite Eq. (\ref{eqmot}) as $i\, \partial_t \varphi_{x,y} = \Omega\left[\varphi_{x+1,y+1} + \varphi_{x+1,y-1} + \varphi_{x-1,y+1} +\varphi_{x-1,y-1}\right].$ We now move to Fourier space, $\varphi_{x,y} = \frac{1}{(2\pi)^2} \int_{-\pi}^\pi dq \int_{-\pi}^\pi dk\, \psi_{k,q}\, e^{i(k x + q y )}$, where $\psi_{k,q} = \sum_{x=0}^\infty \sum_{y=-\infty}^\infty \varphi_{x,y}\, e^{-i(k x + q y )}$.
\begin{equation}
\frac{1}{(2\pi)^2} \int_{-\pi}^\pi dq \int_{-\pi}^\pi dk\, e^{i(k x + q y )} \left[ i\, \partial_t \psi_{k,q} - \Omega \left(\psi_{k,q}\, e^{i(k + q)} + \psi_{k,q}\, e^{i(k - q)}  + \psi_{k,q}\, e^{-i(k-q)}  + \psi_{k,q}\, e^{-i(k+ q)} \right) \right] = 0.
\label{eqmotFour}
\end{equation}
which holds if every Fourier mode satisfies $i\, \partial_t \psi_{k,q} = 4 \Omega \cos q \cos k\, \psi_{k,q}$. We assume the system starts from the initial condition $\varphi_{x,y}(0) = \delta_{x,1} \delta_{y,0}$, or equivalently $\psi_{k,q}(0) = e^{-i k}$. The modes therefore evolve in time as $\psi_{k,q}(t) = e^{-i k} e^{-i 4 \Omega t \cos q \cos k}$, and in real space we obtain
\begin{eqnarray}
\varphi_{x,y}(t) &=& \frac{1}{(2\pi)^2} \int_{-\pi}^\pi\! dq \int_{-\pi}^\pi\! dk\ e^{i\left[k (x-1) + q y - 4\Omega t \cos q \cos k\right]}\nonumber\\ 
&=& \frac{2}{(2\pi)^2} \int_{-\pi}^{\pi}\! dV_{+} e^{i \left[V_+ \left[(x-1) + y\right] - 2\Omega t \cos (2 V_{+})\right]} \int_{-\pi}^{\pi}\! dV_-  e^{i \left[V_-  \left[(x-1) - y\right] - 2\Omega t \cos(2 V_{-})\right]},
\label{Solution}
\end{eqnarray}
where we have changed the integration variables to $V_{\pm} = (k \pm q)/2$ so that the double Fourier integral factorizes. By using the Jacobi-Anger identity $e^{i w \cos \theta} = \sum_{n=-\infty}^{\infty} i^n J_n(w)\, e^{i n \theta}$, where $J_n(w)$ is the $n$-th Bessel function of the first kind, we can compute the integrals:
\begin{equation}
 \int_{-\pi}^{\pi}\! dV_{\pm} e^{i \left[V_\pm \left[(x-1) \pm y\right] - 2\Omega t \cos (2 V_{\pm})\right]}\! =\! \int_{-\pi}^{\pi}\! dV_{\pm}\! \sum_{n=-\infty}^{\infty}\! i^n J_n(-2\Omega t)\, e^{i V_{\pm} \left(2 n + \left[(x-1) \pm y\right]\right)}\! =\! 2\pi i^{-\frac{(x-1) \pm y}{2}} J_{-\frac{(x-1) \pm y}{2}}(-2\Omega t),
\end{equation}
where the orthogonality relation $\int_{-\pi}^{\pi} d\theta\, e^{i (n+m) \theta} = 2\pi \delta_{n,-m}$ has been used. In the basis $|x,X\rangle$, the solution is
\begin{equation}
\varphi_{x,X}(t) = 2  i^{-(x-1)} J_{\frac{x-1}{2}+X}(2\Omega t)\,  J_{\frac{x-1}{2}-X}(2\Omega t).
\label{Solution2}
\end{equation}
We have used the fact that $J_n(w)$ is invariant under a simultaneous change of sign of the index (which is an integer) $n\to-n$ and the (real) argument $w\to-w$. (Indeed, as for domain length $x$ even, the center of mass $X$ is a half integer, and for $x$ odd, $X$ is an integer, the index $(x-1)/2 \pm X$ is always an integer.)

\begin{figure}[h!]
\includegraphics[scale=0.165]{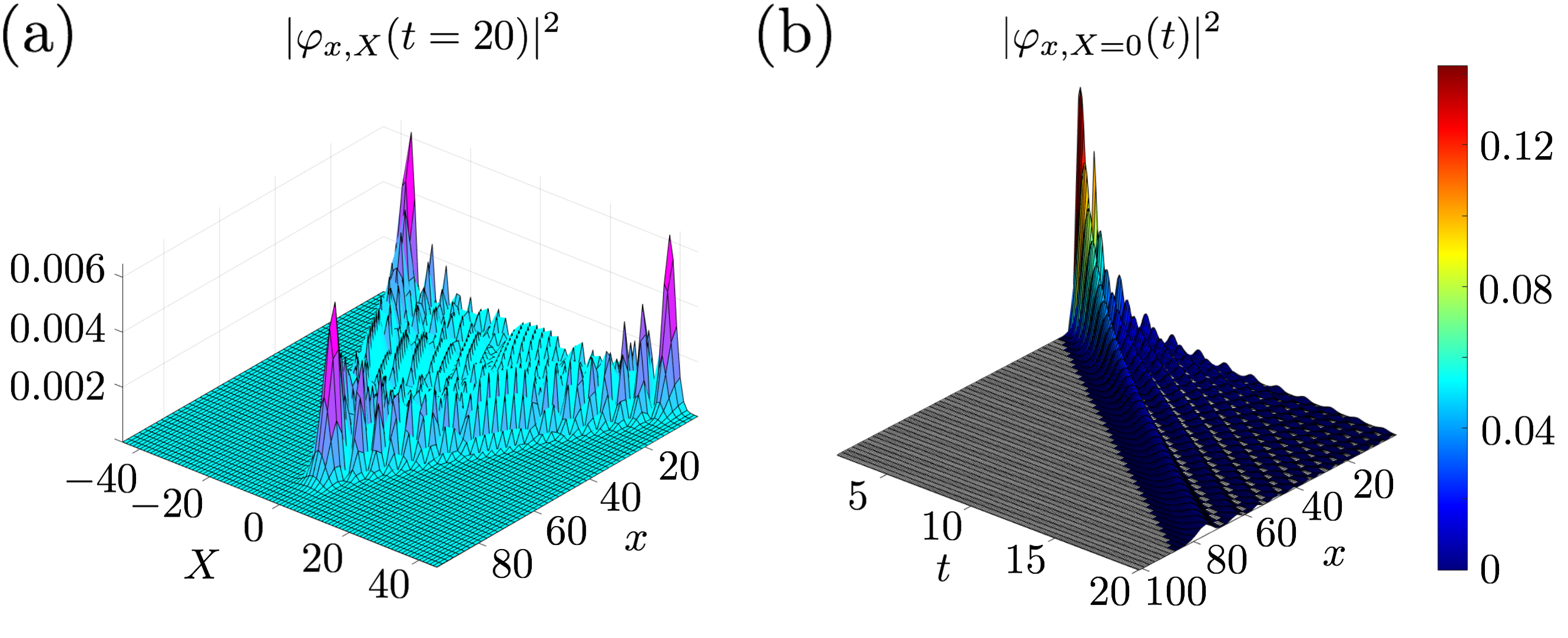}
\caption{ {\sf \bf Solution of the quantum domain-wall model.}
(a) Squared modulus of the solution for $t=20$, $|\varphi_{x,X}(t=20)|^2$  ($\Omega=1$), where three prominent peaks (two of them related by $X\to-X$ symmetry) can be distinguished. (b)  Squared modulus of the solution for $X=0$, $|\varphi_{x,X=0}(t)|^2$  ($\Omega=1$), where the maximum shows a ballistic propagating wavefront $x \approx 4\Omega t + 1$.}
\label{FigHopping}
\end{figure}

We next extract some physical consequences from the solution $\varphi_{x,X}(t)$. The dependence of $|\varphi_{x,X}(t)|^2$ on the center of mass variable is $X$ is shown in Fig. \ref{FigHopping} (a) for $\Omega=1$ and $t=20$ (essentially identical results are obtained for other time points), where the largest values lie along the line segments $\frac{x-1}{2} \pm X \approx 2\Omega t$, for $\frac{x-1}{2}$ in $[0,2\Omega t]$ which roughly corresponds to the position of the maxima of one of the two Bessel functions in Eq. (\ref{Solution2}). The probability amplitude peaks quite prominently around the positions where both Bessel functions align, which are $X\approx 0$ and $x\approx 4\Omega t + 1$, corresponding to a ballistic growth of the domain without center of mass motion, and $X \approx \pm 2 \Omega t$ and $x \approx 1$, which reflects a solitary wave solution without growth. As the latter cannot transform the system, and moreover it is expected to be strongly affected by the presence of decay in realistic settings, we focus on the former in the main text.

In Fig. \ref{FigHopping} (b) we show the modulus squared of the solution for $X=0$, $|\varphi_{x,X=0}(t)|^2$, for a range of $x$ and $t$ ($\Omega=1$). As time increases, the solution is distributed over an ever increasing range of domain size $x$. Moreover, the maximum of the probability amplitude is reached at the position of the propagating wavefront, which follows $x\approx 4\Omega t + 1$.

In the main text we compare the results that arise from this domain-wall model with quantum jump Monte Carlo (QJMC) simulations in the limit of small dephasing rate $\gamma$. In order to do so we calculate the average value of the density $n_k(t)$, which for given $x$ and $X$ is given by the projection operator $\hat{n}_k(x,X) = \Theta(k - [X-\frac{x-1}{2}])\, \Theta([X+\frac{x-1}{2}]-k)$, where $\Theta(\cdot)=0$ if its argument is negative, and $1$ otherwise. For a fixed value of $X$, we obtain
\begin{equation}
\langle n_k(t) \rangle = \sum_{x=0}^\infty \hat{n}_k(x,X) |\varphi_{x,X}(t)|^2 = \sum_{x\geq 2|k-X| + 1} |\varphi_{x,X}(t)|^2.
\end{equation}
which is spatially symmetric around the center of mass $X$. In the main text, we show the time evolution of $\langle n_k(t) \rangle$ for $X=0$ [see Fig. 4 (c) and (d)].

\subsection{Nucleation rate in the weakly dissipative regime}

Nucleation events in the dissipative dynamics are identified with incoherent processes that cause the collapse of an atom onto the excited state. The underlying idea is that the dominant contribution to the generation of spontaneous excitation events that have the strength needed to trigger a growth process is associated with these projection events. When the dephasing rate $\gamma$ is moderate or weak compared to the driving $\Omega$, the domain growth appears to be ballistic and to follow the predictions based on the domain-wall model presented in the previous subsection (see main text). In such regimes the nucleation rate can be estimated by resorting to a quantum jump approach \cite{dalibard1992SM,plenio1998SM}. Starting from a configuration without excitations, in the absence of incoherent processes, the state of a given atom at time $t$ is
$|\psi(t)\rangle = |\psi^{'}(t)\rangle/||\psi^{'}(t)||$, where $|\psi^{'}(t)\rangle = \exp{(-i H_\text{eff}\, t)} |\! \downarrow\rangle$ results from the (non-unitary) time evolution given by the (non-Hermitian) effective Hamiltonian
\begin{equation}
H_\text{eff} = H_{k} - \frac{i \gamma}{2} n = 
  \left( {\begin{array}{cc}
   V- i \gamma/2 & \Omega \\
   \Omega & 0\\
  \end{array} } \right)
\end{equation}
where $H_{k} = V n_k+ \Omega \sigma^x_k$ is the one-body Hamiltonian, and we have taken into account that all sites evolve identically during the initial stages. In a step $\delta t$, the probability of having an incoherent jump to state $|\!\uparrow\rangle$ is 
\begin{equation}
\delta p = \gamma \langle\psi(t)| n  |\psi(t)\rangle \delta t \approx \frac{2 \gamma \Omega^2}{V^2} \left[1 - \cos{(V t)}\right] \delta t,
\end{equation}
where we have expanded to lowest order in $\Omega/V$ and $\gamma/V$, as we consider the (experimentally relevant) case in which 
$V \gg \Omega, \gamma$. This allows us to easily extract an analytical expression for the nucleation rate, but other regimes could be also studied in principle. The average nucleation rate under such strong-interaction conditions is therefore $2 \gamma \Omega^2/V^2$, which is used in the main text to propose an Avrami equation for the system in the limit of weak dephasing.


\begin{thebibliography}{64}%
\makeatletter
\providecommand \@ifxundefined [1]{%
 \@ifx{#1\undefined}
}%
\providecommand \@ifnum [1]{%
 \ifnum #1\expandafter \@firstoftwo
 \else \expandafter \@secondoftwo
 \fi
}%
\providecommand \@ifx [1]{%
 \ifx #1\expandafter \@firstoftwo
 \else \expandafter \@secondoftwo
 \fi
}%
\providecommand \natexlab [1]{#1}%
\providecommand \enquote  [1]{``#1''}%
\providecommand \bibnamefont  [1]{#1}%
\providecommand \bibfnamefont [1]{#1}%
\providecommand \citenamefont [1]{#1}%
\providecommand \href@noop [0]{\@secondoftwo}%
\providecommand \href [0]{\begingroup \@sanitize@url \@href}%
\providecommand \@href[1]{\@@startlink{#1}\@@href}%
\providecommand \@@href[1]{\endgroup#1\@@endlink}%
\providecommand \@sanitize@url [0]{\catcode `\\12\catcode `\$12\catcode
  `\&12\catcode `\#12\catcode `\^12\catcode `\_12\catcode `\%12\relax}%
\providecommand \@@startlink[1]{}%
\providecommand \@@endlink[0]{}%
\providecommand \url  [0]{\begingroup\@sanitize@url \@url }%
\providecommand \@url [1]{\endgroup\@href {#1}{\urlprefix }}%
\providecommand \urlprefix  [0]{URL }%
\providecommand \Eprint [0]{\href }%
\providecommand \doibase [0]{http://dx.doi.org/}%
\providecommand \selectlanguage [0]{\@gobble}%
\providecommand \bibinfo  [0]{\@secondoftwo}%
\providecommand \bibfield  [0]{\@secondoftwo}%
\providecommand \translation [1]{[#1]}%
\providecommand \BibitemOpen [0]{}%
\providecommand \bibitemStop [0]{}%
\providecommand \bibitemNoStop [0]{.\EOS\space}%
\providecommand \EOS [0]{\spacefactor3000\relax}%
\providecommand \BibitemShut  [1]{\csname bibitem#1\endcsname}%
\let\auto@bib@innerbib\@empty
\bibitem [{\citenamefont {Kolmogorov}(1937)}]{kolmogorov1937}%
  \BibitemOpen
  \bibfield  {author} {\bibinfo {author} {\bibfnamefont {A.~E.}\ \bibnamefont
  {Kolmogorov}},\ }\bibfield  {title} {\enquote {\bibinfo {title} {On the
  statistic theory of metal crystallization},}\ }\href@noop {} {\bibfield
  {journal} {\bibinfo  {journal} {Izv. Akad. Nauk. SSSR Ser. Mat.}\ }\textbf
  {\bibinfo {volume} {1}},\ \bibinfo {pages} {333} (\bibinfo {year}
  {1937})}\BibitemShut {NoStop}%
\bibitem [{\citenamefont {Avrami}(1939)}]{avrami1939}%
  \BibitemOpen
  \bibfield  {author} {\bibinfo {author} {\bibfnamefont {M.}~\bibnamefont
  {Avrami}},\ }\bibfield  {title} {\enquote {\bibinfo {title} {Kinetics of
  phase change. {I} {G}eneral theory},}\ }\href@noop {} {\bibfield  {journal}
  {\bibinfo  {journal} {J. Chem. Phys.}\ }\textbf {\bibinfo {volume} {7}},\
  \bibinfo {pages} {1103} (\bibinfo {year} {1939})}\BibitemShut {NoStop}%
\bibitem [{\citenamefont {Johnson}\ and\ \citenamefont
  {Mehl}(1939)}]{johnson1939}%
  \BibitemOpen
  \bibfield  {author} {\bibinfo {author} {\bibfnamefont {W.~A.}\ \bibnamefont
  {Johnson}}\ and\ \bibinfo {author} {\bibfnamefont {R.~F.}\ \bibnamefont
  {Mehl}},\ }\bibfield  {title} {\enquote {\bibinfo {title} {Reaction kinetics
  in processes of nucleation and growth},}\ }\href@noop {} {\bibfield
  {journal} {\bibinfo  {journal} {Trans. Am. Inst. Min. Eng.}\ }\textbf
  {\bibinfo {volume} {135}},\ \bibinfo {pages} {416} (\bibinfo {year}
  {1939})}\BibitemShut {NoStop}%
\bibitem [{\citenamefont {Avrami}(1940)}]{avrami1940}%
  \BibitemOpen
  \bibfield  {author} {\bibinfo {author} {\bibfnamefont {M.}~\bibnamefont
  {Avrami}},\ }\bibfield  {title} {\enquote {\bibinfo {title} {Kinetics of
  phase change. {II} {T}ransformation-time relations for random distribution of
  nuclei},}\ }\href@noop {} {\bibfield  {journal} {\bibinfo  {journal} {J.
  Chem. Phys.}\ }\textbf {\bibinfo {volume} {8}},\ \bibinfo {pages} {212}
  (\bibinfo {year} {1940})}\BibitemShut {NoStop}%
\bibitem [{\citenamefont {Avrami}(1941)}]{avrami1941}%
  \BibitemOpen
  \bibfield  {author} {\bibinfo {author} {\bibfnamefont {M.}~\bibnamefont
  {Avrami}},\ }\bibfield  {title} {\enquote {\bibinfo {title} {Granulation,
  phase change, and microstructure kinetics of phase change. {III}},}\
  }\href@noop {} {\bibfield  {journal} {\bibinfo  {journal} {J. Chem. Phys.}\
  }\textbf {\bibinfo {volume} {9}},\ \bibinfo {pages} {177} (\bibinfo {year}
  {1941})}\BibitemShut {NoStop}%
\bibitem [{\citenamefont {Christian}(2002)}]{christian2002}%
  \BibitemOpen
  \bibfield  {author} {\bibinfo {author} {\bibfnamefont {J.~W.}\ \bibnamefont
  {Christian}},\ }\href@noop {} {\emph {\bibinfo {title} {The Theory of
  Transformations in Metals and Alloys}}}\ (\bibinfo  {publisher} {Pergamon,
  Oxford},\ \bibinfo {year} {2002})\BibitemShut {NoStop}%
\bibitem [{\citenamefont {Jun}\ \emph {et~al.}(2005)\citenamefont {Jun},
  \citenamefont {Zhang},\ and\ \citenamefont {Bechhoefer}}]{jun2005a}%
  \BibitemOpen
  \bibfield  {author} {\bibinfo {author} {\bibfnamefont {S.}~\bibnamefont
  {Jun}}, \bibinfo {author} {\bibfnamefont {H.}~\bibnamefont {Zhang}}, \ and\
  \bibinfo {author} {\bibfnamefont {J.}~\bibnamefont {Bechhoefer}},\ }\bibfield
   {title} {\enquote {\bibinfo {title} {Nucleation and growth in one dimension.
  {I}. {T}he generalized {K}olmogorov-{J}ohnson-{M}ehl-{A}vrami model},}\
  }\href@noop {} {\bibfield  {journal} {\bibinfo  {journal} {Phys. Rev. E}\
  }\textbf {\bibinfo {volume} {71}},\ \bibinfo {pages} {011908} (\bibinfo
  {year} {2005})}\BibitemShut {NoStop}%
\bibitem [{\citenamefont {Jun}\ and\ \citenamefont
  {Bechhoefer}(2005)}]{jun2005b}%
  \BibitemOpen
  \bibfield  {author} {\bibinfo {author} {\bibfnamefont {S.}~\bibnamefont
  {Jun}}\ and\ \bibinfo {author} {\bibfnamefont {J.}~\bibnamefont
  {Bechhoefer}},\ }\bibfield  {title} {\enquote {\bibinfo {title} {Nucleation
  and growth in one dimension. {II}. {A}pplication to {D}{N}{A} replication
  kinetics},}\ }\href@noop {} {\bibfield  {journal} {\bibinfo  {journal} {Phys.
  Rev. E}\ }\textbf {\bibinfo {volume} {71}},\ \bibinfo {pages} {011909}
  (\bibinfo {year} {2005})}\BibitemShut {NoStop}%
\bibitem [{\citenamefont {Avramov}(2007)}]{avramov2007}%
  \BibitemOpen
  \bibfield  {author} {\bibinfo {author} {\bibfnamefont {I.}~\bibnamefont
  {Avramov}},\ }\bibfield  {title} {\enquote {\bibinfo {title} {Kinetics of
  distribution of infections in networks},}\ }\href@noop {} {\bibfield
  {journal} {\bibinfo  {journal} {Physica A}\ }\textbf {\bibinfo {volume}
  {379}},\ \bibinfo {pages} {615} (\bibinfo {year} {2007})}\BibitemShut
  {NoStop}%
\bibitem [{\citenamefont {Guti{\'e}rrez}\ and\ \citenamefont
  {Garrahan}(2016)}]{gutierrez2016a}%
  \BibitemOpen
  \bibfield  {author} {\bibinfo {author} {\bibfnamefont {R.}~\bibnamefont
  {Guti{\'e}rrez}}\ and\ \bibinfo {author} {\bibfnamefont {J.~P.}\ \bibnamefont
  {Garrahan}},\ }\bibfield  {title} {\enquote {\bibinfo {title} {Front
  propagation versus bulk relaxation in the annealing dynamics of a kinetically
  constrained model of ultrastable glasses},}\ }\href@noop {} {\bibfield
  {journal} {\bibinfo  {journal} {J. Stat. Mech.: Theor. Exp.}\ }\textbf
  {\bibinfo {volume} {2016}},\ \bibinfo {pages} {074005} (\bibinfo {year}
  {2016})}\BibitemShut {NoStop}%
\bibitem [{\citenamefont {Jack}\ and\ \citenamefont
  {Berthier}(2016)}]{jack2016}%
  \BibitemOpen
  \bibfield  {author} {\bibinfo {author} {\bibfnamefont {R.~L.}\ \bibnamefont
  {Jack}}\ and\ \bibinfo {author} {\bibfnamefont {L.}~\bibnamefont
  {Berthier}},\ }\bibfield  {title} {\enquote {\bibinfo {title} {The melting of
  stable glasses is governed by nucleation-and-growth dynamics},}\ }\href@noop
  {} {\bibfield  {journal} {\bibinfo  {journal} {J. Chem. Phys.}\ }\textbf
  {\bibinfo {volume} {144}},\ \bibinfo {pages} {244506} (\bibinfo {year}
  {2016})}\BibitemShut {NoStop}%
\bibitem [{\citenamefont {Rigol}\ \emph {et~al.}(2006)\citenamefont {Rigol},
  \citenamefont {Muramatsu},\ and\ \citenamefont {Olshanii}}]{rigol2006}%
  \BibitemOpen
  \bibfield  {author} {\bibinfo {author} {\bibfnamefont {M.}~\bibnamefont
  {Rigol}}, \bibinfo {author} {\bibfnamefont {A.}~\bibnamefont {Muramatsu}}, \
  and\ \bibinfo {author} {\bibfnamefont {M.}~\bibnamefont {Olshanii}},\
  }\bibfield  {title} {\enquote {\bibinfo {title} {Hard-core bosons on optical
  superlattices: Dynamics and relaxation in the superfluid and insulating
  regimes},}\ }\href@noop {} {\bibfield  {journal} {\bibinfo  {journal} {Phys.
  Rev. A}\ }\textbf {\bibinfo {volume} {74}},\ \bibinfo {pages} {053616}
  (\bibinfo {year} {2006})}\BibitemShut {NoStop}%
\bibitem [{\citenamefont {Cazalilla}(2006)}]{cazalilla2006}%
  \BibitemOpen
  \bibfield  {author} {\bibinfo {author} {\bibfnamefont {M.~A.}\ \bibnamefont
  {Cazalilla}},\ }\bibfield  {title} {\enquote {\bibinfo {title} {Effect of
  suddenly turning on interactions in the {L}uttinger model},}\ }\href@noop {}
  {\bibfield  {journal} {\bibinfo  {journal} {Phys. Rev. Lett.}\ }\textbf
  {\bibinfo {volume} {97}},\ \bibinfo {pages} {156403} (\bibinfo {year}
  {2006})}\BibitemShut {NoStop}%
\bibitem [{\citenamefont {Calabrese}\ and\ \citenamefont
  {Cardy}(2006)}]{calabrese2006}%
  \BibitemOpen
  \bibfield  {author} {\bibinfo {author} {\bibfnamefont {P.}~\bibnamefont
  {Calabrese}}\ and\ \bibinfo {author} {\bibfnamefont {J.}~\bibnamefont
  {Cardy}},\ }\bibfield  {title} {\enquote {\bibinfo {title} {Time dependence
  of correlation functions following a quantum quench},}\ }\href@noop {}
  {\bibfield  {journal} {\bibinfo  {journal} {Phys. Rev. Lett.}\ }\textbf
  {\bibinfo {volume} {96}},\ \bibinfo {pages} {136801} (\bibinfo {year}
  {2006})}\BibitemShut {NoStop}%
\bibitem [{\citenamefont {Kollath}\ \emph {et~al.}(2007)\citenamefont
  {Kollath}, \citenamefont {L{\"a}uchli},\ and\ \citenamefont
  {Altman}}]{kollath2007}%
  \BibitemOpen
  \bibfield  {author} {\bibinfo {author} {\bibfnamefont {C.}~\bibnamefont
  {Kollath}}, \bibinfo {author} {\bibfnamefont {A.~M.}\ \bibnamefont
  {L{\"a}uchli}}, \ and\ \bibinfo {author} {\bibfnamefont {E.}~\bibnamefont
  {Altman}},\ }\bibfield  {title} {\enquote {\bibinfo {title} {Quench dynamics
  and nonequilibrium phase diagram of the {B}ose-{H}ubbard model},}\
  }\href@noop {} {\bibfield  {journal} {\bibinfo  {journal} {Phys. Rev. Lett.}\
  }\textbf {\bibinfo {volume} {98}},\ \bibinfo {pages} {180601} (\bibinfo
  {year} {2007})}\BibitemShut {NoStop}%
\bibitem [{\citenamefont {L{\"a}uchli}\ and\ \citenamefont
  {Kollath}(2008)}]{lauchli2008}%
  \BibitemOpen
  \bibfield  {author} {\bibinfo {author} {\bibfnamefont {A.~M.}\ \bibnamefont
  {L{\"a}uchli}}\ and\ \bibinfo {author} {\bibfnamefont {C.}~\bibnamefont
  {Kollath}},\ }\bibfield  {title} {\enquote {\bibinfo {title} {Spreading of
  correlations and entanglement after a quench in the one-dimensional
  {B}ose--{H}ubbard model},}\ }\href@noop {} {\bibfield  {journal} {\bibinfo
  {journal} {J. Stat. Mech.: Theor. Exp.}\ }\textbf {\bibinfo {volume}
  {2008}},\ \bibinfo {pages} {P05018} (\bibinfo {year} {2008})}\BibitemShut
  {NoStop}%
\bibitem [{\citenamefont {Mitra}(2017)}]{mitra2017}%
  \BibitemOpen
  \bibfield  {author} {\bibinfo {author} {\bibfnamefont {A.}~\bibnamefont
  {Mitra}},\ }\bibfield  {title} {\enquote {\bibinfo {title} {Quantum quench
  dynamics},}\ }\href@noop {} {\bibfield  {journal} {\bibinfo  {journal}
  {arXiv:1703.09740}\ } (\bibinfo {year} {2017})}\BibitemShut {NoStop}%
\bibitem [{\citenamefont {Kinoshita}\ \emph {et~al.}(2006)\citenamefont
  {Kinoshita}, \citenamefont {Wenger},\ and\ \citenamefont
  {Weiss}}]{kinoshita2006}%
  \BibitemOpen
  \bibfield  {author} {\bibinfo {author} {\bibfnamefont {T.}~\bibnamefont
  {Kinoshita}}, \bibinfo {author} {\bibfnamefont {T.}~\bibnamefont {Wenger}}, \
  and\ \bibinfo {author} {\bibfnamefont {D.~S.}\ \bibnamefont {Weiss}},\
  }\bibfield  {title} {\enquote {\bibinfo {title} {A quantum newton's
  cradle},}\ }\href@noop {} {\bibfield  {journal} {\bibinfo  {journal}
  {Nature}\ }\textbf {\bibinfo {volume} {440}},\ \bibinfo {pages} {900}
  (\bibinfo {year} {2006})}\BibitemShut {NoStop}%
\bibitem [{\citenamefont {Hofferberth}\ \emph {et~al.}(2007)\citenamefont
  {Hofferberth}, \citenamefont {Lesanovsky}, \citenamefont {Fischer},
  \citenamefont {Schumm},\ and\ \citenamefont
  {Schmiedmayer}}]{hofferberth2007}%
  \BibitemOpen
  \bibfield  {author} {\bibinfo {author} {\bibfnamefont {S.}~\bibnamefont
  {Hofferberth}}, \bibinfo {author} {\bibfnamefont {I.}~\bibnamefont
  {Lesanovsky}}, \bibinfo {author} {\bibfnamefont {B.}~\bibnamefont {Fischer}},
  \bibinfo {author} {\bibfnamefont {T.}~\bibnamefont {Schumm}}, \ and\ \bibinfo
  {author} {\bibfnamefont {J.}~\bibnamefont {Schmiedmayer}},\ }\bibfield
  {title} {\enquote {\bibinfo {title} {Non-equilibrium coherence dynamics in
  one-dimensional bose gases},}\ }\href@noop {} {\bibfield  {journal} {\bibinfo
   {journal} {Nature}\ }\textbf {\bibinfo {volume} {449}},\ \bibinfo {pages}
  {324} (\bibinfo {year} {2007})}\BibitemShut {NoStop}%
\bibitem [{\citenamefont {Chen}\ \emph {et~al.}(2011)\citenamefont {Chen},
  \citenamefont {White}, \citenamefont {Borries},\ and\ \citenamefont
  {DeMarco}}]{chen2011}%
  \BibitemOpen
  \bibfield  {author} {\bibinfo {author} {\bibfnamefont {D.}~\bibnamefont
  {Chen}}, \bibinfo {author} {\bibfnamefont {M.}~\bibnamefont {White}},
  \bibinfo {author} {\bibfnamefont {C.}~\bibnamefont {Borries}}, \ and\
  \bibinfo {author} {\bibfnamefont {B.}~\bibnamefont {DeMarco}},\ }\bibfield
  {title} {\enquote {\bibinfo {title} {Quantum quench of an atomic {M}ott
  insulator},}\ }\href@noop {} {\bibfield  {journal} {\bibinfo  {journal}
  {Phys. Rev. Lett.}\ }\textbf {\bibinfo {volume} {106}},\ \bibinfo {pages}
  {235304} (\bibinfo {year} {2011})}\BibitemShut {NoStop}%
\bibitem [{\citenamefont {Gring}\ \emph {et~al.}(2012)\citenamefont {Gring},
  \citenamefont {Kuhnert}, \citenamefont {Langen}, \citenamefont {Kitagawa},
  \citenamefont {Rauer}, \citenamefont {Schreitl}, \citenamefont {Mazets},
  \citenamefont {Smith}, \citenamefont {Demler},\ and\ \citenamefont
  {Schmiedmayer}}]{gring2012}%
  \BibitemOpen
  \bibfield  {author} {\bibinfo {author} {\bibfnamefont {M.}~\bibnamefont
  {Gring}}, \bibinfo {author} {\bibfnamefont {M.}~\bibnamefont {Kuhnert}},
  \bibinfo {author} {\bibfnamefont {T.}~\bibnamefont {Langen}}, \bibinfo
  {author} {\bibfnamefont {T.}~\bibnamefont {Kitagawa}}, \bibinfo {author}
  {\bibfnamefont {B.}~\bibnamefont {Rauer}}, \bibinfo {author} {\bibfnamefont
  {M.}~\bibnamefont {Schreitl}}, \bibinfo {author} {\bibfnamefont
  {I.}~\bibnamefont {Mazets}}, \bibinfo {author} {\bibfnamefont {D.~A.}\
  \bibnamefont {Smith}}, \bibinfo {author} {\bibfnamefont {E.}~\bibnamefont
  {Demler}}, \ and\ \bibinfo {author} {\bibfnamefont {J.}~\bibnamefont
  {Schmiedmayer}},\ }\bibfield  {title} {\enquote {\bibinfo {title} {Relaxation
  and prethermalization in an isolated quantum system},}\ }\href@noop {}
  {\bibfield  {journal} {\bibinfo  {journal} {Science}\ }\textbf {\bibinfo
  {volume} {337}},\ \bibinfo {pages} {1318} (\bibinfo {year}
  {2012})}\BibitemShut {NoStop}%
\bibitem [{\citenamefont {Schreiber}\ \emph {et~al.}(2015)\citenamefont
  {Schreiber}, \citenamefont {Hodgman}, \citenamefont {Bordia}, \citenamefont
  {L{\"u}schen}, \citenamefont {Fischer}, \citenamefont {Vosk}, \citenamefont
  {Altman}, \citenamefont {Schneider},\ and\ \citenamefont
  {Bloch}}]{schreiber2015}%
  \BibitemOpen
  \bibfield  {author} {\bibinfo {author} {\bibfnamefont {M.}~\bibnamefont
  {Schreiber}}, \bibinfo {author} {\bibfnamefont {S.~S.}\ \bibnamefont
  {Hodgman}}, \bibinfo {author} {\bibfnamefont {P.}~\bibnamefont {Bordia}},
  \bibinfo {author} {\bibfnamefont {H.~P.}\ \bibnamefont {L{\"u}schen}},
  \bibinfo {author} {\bibfnamefont {M.~H.}\ \bibnamefont {Fischer}}, \bibinfo
  {author} {\bibfnamefont {R.}~\bibnamefont {Vosk}}, \bibinfo {author}
  {\bibfnamefont {E.}~\bibnamefont {Altman}}, \bibinfo {author} {\bibfnamefont
  {U.}~\bibnamefont {Schneider}}, \ and\ \bibinfo {author} {\bibfnamefont
  {I.}~\bibnamefont {Bloch}},\ }\bibfield  {title} {\enquote {\bibinfo {title}
  {Observation of many-body localization of interacting fermions in a
  quasirandom optical lattice},}\ }\href@noop {} {\bibfield  {journal}
  {\bibinfo  {journal} {Science}\ }\textbf {\bibinfo {volume} {349}},\ \bibinfo
  {pages} {842} (\bibinfo {year} {2015})}\BibitemShut {NoStop}%
\bibitem [{\citenamefont {Kaufman}\ \emph {et~al.}(2016)\citenamefont
  {Kaufman}, \citenamefont {Tai}, \citenamefont {Lukin}, \citenamefont
  {Rispoli}, \citenamefont {Schittko}, \citenamefont {Preiss},\ and\
  \citenamefont {Greiner}}]{kaufman2016}%
  \BibitemOpen
  \bibfield  {author} {\bibinfo {author} {\bibfnamefont {A.~M.}\ \bibnamefont
  {Kaufman}}, \bibinfo {author} {\bibfnamefont {M.~E.}\ \bibnamefont {Tai}},
  \bibinfo {author} {\bibfnamefont {A.}~\bibnamefont {Lukin}}, \bibinfo
  {author} {\bibfnamefont {M.}~\bibnamefont {Rispoli}}, \bibinfo {author}
  {\bibfnamefont {R.}~\bibnamefont {Schittko}}, \bibinfo {author}
  {\bibfnamefont {P.~M.}\ \bibnamefont {Preiss}}, \ and\ \bibinfo {author}
  {\bibfnamefont {M.}~\bibnamefont {Greiner}},\ }\bibfield  {title} {\enquote
  {\bibinfo {title} {Quantum thermalization through entanglement in an isolated
  many-body system},}\ }\href@noop {} {\bibfield  {journal} {\bibinfo
  {journal} {Science}\ }\textbf {\bibinfo {volume} {353}},\ \bibinfo {pages}
  {794} (\bibinfo {year} {2016})}\BibitemShut {NoStop}%
\bibitem [{\citenamefont {Schachenmayer}\ \emph {et~al.}(2013)\citenamefont
  {Schachenmayer}, \citenamefont {Lanyon}, \citenamefont {Roos},\ and\
  \citenamefont {Daley}}]{schachenmayer2013}%
  \BibitemOpen
  \bibfield  {author} {\bibinfo {author} {\bibfnamefont {J.}~\bibnamefont
  {Schachenmayer}}, \bibinfo {author} {\bibfnamefont {B.~P.}\ \bibnamefont
  {Lanyon}}, \bibinfo {author} {\bibfnamefont {C.~F.}\ \bibnamefont {Roos}}, \
  and\ \bibinfo {author} {\bibfnamefont {A.~J.}\ \bibnamefont {Daley}},\
  }\bibfield  {title} {\enquote {\bibinfo {title} {Entanglement growth in
  quench dynamics with variable range interactions},}\ }\href@noop {}
  {\bibfield  {journal} {\bibinfo  {journal} {Phys. Rev. X}\ }\textbf {\bibinfo
  {volume} {3}},\ \bibinfo {pages} {031015} (\bibinfo {year}
  {2013})}\BibitemShut {NoStop}%
\bibitem [{\citenamefont {Richerme}\ \emph {et~al.}(2014)\citenamefont
  {Richerme}, \citenamefont {Gong}, \citenamefont {Lee}, \citenamefont {Senko},
  \citenamefont {Smith}, \citenamefont {Foss-Feig}, \citenamefont {Michalakis},
  \citenamefont {Gorshkov},\ and\ \citenamefont {Monroe}}]{richerme2014}%
  \BibitemOpen
  \bibfield  {author} {\bibinfo {author} {\bibfnamefont {P.}~\bibnamefont
  {Richerme}}, \bibinfo {author} {\bibfnamefont {Z.-X.}\ \bibnamefont {Gong}},
  \bibinfo {author} {\bibfnamefont {A.}~\bibnamefont {Lee}}, \bibinfo {author}
  {\bibfnamefont {C.}~\bibnamefont {Senko}}, \bibinfo {author} {\bibfnamefont
  {J.}~\bibnamefont {Smith}}, \bibinfo {author} {\bibfnamefont
  {M.}~\bibnamefont {Foss-Feig}}, \bibinfo {author} {\bibfnamefont
  {S.}~\bibnamefont {Michalakis}}, \bibinfo {author} {\bibfnamefont {A.~V.}\
  \bibnamefont {Gorshkov}}, \ and\ \bibinfo {author} {\bibfnamefont
  {C.}~\bibnamefont {Monroe}},\ }\bibfield  {title} {\enquote {\bibinfo {title}
  {Non-local propagation of correlations in long-range interacting quantum
  systems},}\ }\href@noop {} {\bibfield  {journal} {\bibinfo  {journal}
  {Nature}\ }\textbf {\bibinfo {volume} {511}},\ \bibinfo {pages} {198}
  (\bibinfo {year} {2014})}\BibitemShut {NoStop}%
\bibitem [{\citenamefont {Kashuba}\ \emph {et~al.}(2013)\citenamefont
  {Kashuba}, \citenamefont {Kennes}, \citenamefont {Pletyukhov}, \citenamefont
  {Meden},\ and\ \citenamefont {Schoeller}}]{kashuba2013}%
  \BibitemOpen
  \bibfield  {author} {\bibinfo {author} {\bibfnamefont {O.}~\bibnamefont
  {Kashuba}}, \bibinfo {author} {\bibfnamefont {D.~M.}\ \bibnamefont {Kennes}},
  \bibinfo {author} {\bibfnamefont {M.}~\bibnamefont {Pletyukhov}}, \bibinfo
  {author} {\bibfnamefont {V.}~\bibnamefont {Meden}}, \ and\ \bibinfo {author}
  {\bibfnamefont {H.}~\bibnamefont {Schoeller}},\ }\bibfield  {title} {\enquote
  {\bibinfo {title} {Quench dynamics of a dissipative quantum system: {A}
  renormalization group study},}\ }\href@noop {} {\bibfield  {journal}
  {\bibinfo  {journal} {Phys. Rev. B}\ }\textbf {\bibinfo {volume} {88}},\
  \bibinfo {pages} {165133} (\bibinfo {year} {2013})}\BibitemShut {NoStop}%
\bibitem [{\citenamefont {Kennes}\ \emph {et~al.}(2013)\citenamefont {Kennes},
  \citenamefont {Kashuba},\ and\ \citenamefont {Meden}}]{kennes2013}%
  \BibitemOpen
  \bibfield  {author} {\bibinfo {author} {\bibfnamefont {D.~M.}\ \bibnamefont
  {Kennes}}, \bibinfo {author} {\bibfnamefont {O.}~\bibnamefont {Kashuba}}, \
  and\ \bibinfo {author} {\bibfnamefont {V.}~\bibnamefont {Meden}},\ }\bibfield
   {title} {\enquote {\bibinfo {title} {Dynamical regimes of dissipative
  quantum systems},}\ }\href@noop {} {\bibfield  {journal} {\bibinfo  {journal}
  {Phys. Rev. B}\ }\textbf {\bibinfo {volume} {88}},\ \bibinfo {pages} {241110}
  (\bibinfo {year} {2013})}\BibitemShut {NoStop}%
\bibitem [{\citenamefont {Creatore}\ \emph {et~al.}(2014)\citenamefont
  {Creatore}, \citenamefont {Fazio}, \citenamefont {Keeling},\ and\
  \citenamefont {T{\"u}reci}}]{creatore2014}%
  \BibitemOpen
  \bibfield  {author} {\bibinfo {author} {\bibfnamefont {C}~\bibnamefont
  {Creatore}}, \bibinfo {author} {\bibfnamefont {R}~\bibnamefont {Fazio}},
  \bibinfo {author} {\bibfnamefont {J}~\bibnamefont {Keeling}}, \ and\ \bibinfo
  {author} {\bibfnamefont {HE}~\bibnamefont {T{\"u}reci}},\ }\bibfield  {title}
  {\enquote {\bibinfo {title} {Quench dynamics of a disordered array of
  dissipative coupled cavities},}\ }\bibfield  {booktitle} {\emph {\bibinfo
  {booktitle} {Proc. R. Soc. A}},\ }\href@noop {} {\ \textbf {\bibinfo {volume}
  {470}},\ \bibinfo {pages} {20140328} (\bibinfo {year} {2014})}\BibitemShut
  {NoStop}%
\bibitem [{\citenamefont {Henriet}\ and\ \citenamefont
  {Le~Hur}(2016)}]{henriet2016}%
  \BibitemOpen
  \bibfield  {author} {\bibinfo {author} {\bibfnamefont {L.}~\bibnamefont
  {Henriet}}\ and\ \bibinfo {author} {\bibfnamefont {K.}~\bibnamefont
  {Le~Hur}},\ }\bibfield  {title} {\enquote {\bibinfo {title} {Quantum sweeps,
  synchronization, and {K}ibble-{Z}urek physics in dissipative quantum spin
  systems},}\ }\href@noop {} {\bibfield  {journal} {\bibinfo  {journal} {Phys.
  Rev. B}\ }\textbf {\bibinfo {volume} {93}},\ \bibinfo {pages} {064411}
  (\bibinfo {year} {2016})}\BibitemShut {NoStop}%
\bibitem [{\citenamefont {Shapourian}(2016)}]{shapourian2016}%
  \BibitemOpen
  \bibfield  {author} {\bibinfo {author} {\bibfnamefont {H.}~\bibnamefont
  {Shapourian}},\ }\bibfield  {title} {\enquote {\bibinfo {title} {Dynamical
  renormalization-group approach to the spin-boson model},}\ }\href@noop {}
  {\bibfield  {journal} {\bibinfo  {journal} {Phys. Rev. A}\ }\textbf {\bibinfo
  {volume} {93}},\ \bibinfo {pages} {032119} (\bibinfo {year}
  {2016})}\BibitemShut {NoStop}%
\bibitem [{\citenamefont {Bernier}\ \emph {et~al.}(2018)\citenamefont
  {Bernier}, \citenamefont {Tan}, \citenamefont {Bonnes}, \citenamefont {Guo},
  \citenamefont {Poletti},\ and\ \citenamefont {Kollath}}]{bernier2017}%
  \BibitemOpen
  \bibfield  {author} {\bibinfo {author} {\bibfnamefont {J.-S.}\ \bibnamefont
  {Bernier}}, \bibinfo {author} {\bibfnamefont {R.}~\bibnamefont {Tan}},
  \bibinfo {author} {\bibfnamefont {L.}~\bibnamefont {Bonnes}}, \bibinfo
  {author} {\bibfnamefont {C.}~\bibnamefont {Guo}}, \bibinfo {author}
  {\bibfnamefont {D.}~\bibnamefont {Poletti}}, \ and\ \bibinfo {author}
  {\bibfnamefont {C.}~\bibnamefont {Kollath}},\ }\bibfield  {title} {\enquote
  {\bibinfo {title} {Light-cone and diffusive propagation of correlations in a
  many-body dissipative system},}\ }\href@noop {} {\bibfield  {journal}
  {\bibinfo  {journal} {Physical Review Letters}\ }\textbf {\bibinfo {volume}
  {120}},\ \bibinfo {pages} {020401} (\bibinfo {year} {2018})}\BibitemShut
  {NoStop}%
\bibitem [{\citenamefont {Gallagher}(2005)}]{gallagher2005}%
  \BibitemOpen
  \bibfield  {author} {\bibinfo {author} {\bibfnamefont {T.~F.}\ \bibnamefont
  {Gallagher}},\ }\href@noop {} {\emph {\bibinfo {title} {Rydberg atoms}}},\
  Vol.~\bibinfo {volume} {3}\ (\bibinfo  {publisher} {Cambridge University
  Press},\ \bibinfo {year} {2005})\BibitemShut {NoStop}%
\bibitem [{\citenamefont {L{\"o}w}\ \emph {et~al.}(2012)\citenamefont
  {L{\"o}w}, \citenamefont {Weimer}, \citenamefont {Nipper}, \citenamefont
  {Balewski}, \citenamefont {Butscher}, \citenamefont {B{\"u}chler},\ and\
  \citenamefont {Pfau}}]{low2012}%
  \BibitemOpen
  \bibfield  {author} {\bibinfo {author} {\bibfnamefont {R.}~\bibnamefont
  {L{\"o}w}}, \bibinfo {author} {\bibfnamefont {H.}~\bibnamefont {Weimer}},
  \bibinfo {author} {\bibfnamefont {J.}~\bibnamefont {Nipper}}, \bibinfo
  {author} {\bibfnamefont {J.~B.}\ \bibnamefont {Balewski}}, \bibinfo {author}
  {\bibfnamefont {B.}~\bibnamefont {Butscher}}, \bibinfo {author}
  {\bibfnamefont {H.~P.}\ \bibnamefont {B{\"u}chler}}, \ and\ \bibinfo {author}
  {\bibfnamefont {T.}~\bibnamefont {Pfau}},\ }\bibfield  {title} {\enquote
  {\bibinfo {title} {An experimental and theoretical guide to strongly
  interacting {R}ydberg gases},}\ }\href@noop {} {\bibfield  {journal}
  {\bibinfo  {journal} {J. Phys. B: At. Mol. Opt. Phys.}\ }\textbf {\bibinfo
  {volume} {45}},\ \bibinfo {pages} {113001} (\bibinfo {year}
  {2012})}\BibitemShut {NoStop}%
\bibitem [{\citenamefont {Amthor}\ \emph {et~al.}(2010)\citenamefont {Amthor},
  \citenamefont {Giese}, \citenamefont {Hofmann},\ and\ \citenamefont
  {Weidem{\"u}ller}}]{amthor2010}%
  \BibitemOpen
  \bibfield  {author} {\bibinfo {author} {\bibfnamefont {T.}~\bibnamefont
  {Amthor}}, \bibinfo {author} {\bibfnamefont {C.}~\bibnamefont {Giese}},
  \bibinfo {author} {\bibfnamefont {C.~S.}\ \bibnamefont {Hofmann}}, \ and\
  \bibinfo {author} {\bibfnamefont {M.}~\bibnamefont {Weidem{\"u}ller}},\
  }\bibfield  {title} {\enquote {\bibinfo {title} {Evidence of antiblockade in
  an ultracold rydberg gas},}\ }\href@noop {} {\bibfield  {journal} {\bibinfo
  {journal} {Phys. Rev. Lett.}\ }\textbf {\bibinfo {volume} {104}},\ \bibinfo
  {pages} {013001} (\bibinfo {year} {2010})}\BibitemShut {NoStop}%
\bibitem [{\citenamefont {Lesanovsky}\ and\ \citenamefont
  {Garrahan}(2014)}]{lesanovsky2014}%
  \BibitemOpen
  \bibfield  {author} {\bibinfo {author} {\bibfnamefont {I.}~\bibnamefont
  {Lesanovsky}}\ and\ \bibinfo {author} {\bibfnamefont {J.~P.}\ \bibnamefont
  {Garrahan}},\ }\bibfield  {title} {\enquote {\bibinfo {title}
  {Out-of-equilibrium structures in strongly interacting {R}ydberg gases with
  dissipation},}\ }\href@noop {} {\bibfield  {journal} {\bibinfo  {journal}
  {Phys. Rev. A}\ }\textbf {\bibinfo {volume} {90}},\ \bibinfo {pages} {011603}
  (\bibinfo {year} {2014})}\BibitemShut {NoStop}%
\bibitem [{\citenamefont {Marcuzzi}\ \emph {et~al.}(2017)\citenamefont
  {Marcuzzi}, \citenamefont {Min{\'a}{\v{r}}}, \citenamefont {Barredo},
  \citenamefont {de~L{\'e}s{\'e}leuc}, \citenamefont {Labuhn}, \citenamefont
  {Lahaye}, \citenamefont {Browaeys}, \citenamefont {Levi},\ and\ \citenamefont
  {Lesanovsky}}]{marcuzzi2017}%
  \BibitemOpen
  \bibfield  {author} {\bibinfo {author} {\bibfnamefont {M.}~\bibnamefont
  {Marcuzzi}}, \bibinfo {author} {\bibfnamefont {J.}~\bibnamefont
  {Min{\'a}{\v{r}}}}, \bibinfo {author} {\bibfnamefont {D.}~\bibnamefont
  {Barredo}}, \bibinfo {author} {\bibfnamefont {S.}~\bibnamefont
  {de~L{\'e}s{\'e}leuc}}, \bibinfo {author} {\bibfnamefont {H.}~\bibnamefont
  {Labuhn}}, \bibinfo {author} {\bibfnamefont {T.}~\bibnamefont {Lahaye}},
  \bibinfo {author} {\bibfnamefont {A.}~\bibnamefont {Browaeys}}, \bibinfo
  {author} {\bibfnamefont {E.}~\bibnamefont {Levi}}, \ and\ \bibinfo {author}
  {\bibfnamefont {I.}~\bibnamefont {Lesanovsky}},\ }\bibfield  {title}
  {\enquote {\bibinfo {title} {Facilitation dynamics and localization phenomena
  in rydberg lattice gases with position disorder},}\ }\href@noop {} {\bibfield
   {journal} {\bibinfo  {journal} {Phys. Rev. Lett.}\ }\textbf {\bibinfo
  {volume} {118}},\ \bibinfo {pages} {063606} (\bibinfo {year}
  {2017})}\BibitemShut {NoStop}%
\bibitem [{\citenamefont {Simonelli}\ \emph {et~al.}(2016)\citenamefont
  {Simonelli}, \citenamefont {Valado}, \citenamefont {Masella}, \citenamefont
  {Asteria}, \citenamefont {Arimondo}, \citenamefont {Ciampini},\ and\
  \citenamefont {Morsch}}]{simonelli2016}%
  \BibitemOpen
  \bibfield  {author} {\bibinfo {author} {\bibfnamefont {C.}~\bibnamefont
  {Simonelli}}, \bibinfo {author} {\bibfnamefont {M.~M.}\ \bibnamefont
  {Valado}}, \bibinfo {author} {\bibfnamefont {G.}~\bibnamefont {Masella}},
  \bibinfo {author} {\bibfnamefont {L.}~\bibnamefont {Asteria}}, \bibinfo
  {author} {\bibfnamefont {E.}~\bibnamefont {Arimondo}}, \bibinfo {author}
  {\bibfnamefont {D.}~\bibnamefont {Ciampini}}, \ and\ \bibinfo {author}
  {\bibfnamefont {O.}~\bibnamefont {Morsch}},\ }\bibfield  {title} {\enquote
  {\bibinfo {title} {Seeded excitation avalanches in off-resonantly driven
  rydberg gases},}\ }\href@noop {} {\bibfield  {journal} {\bibinfo  {journal}
  {J. Phys. B: At. Mol. Opt. Phys.}\ }\textbf {\bibinfo {volume} {49}},\
  \bibinfo {pages} {154002} (\bibinfo {year} {2016})}\BibitemShut {NoStop}%
\bibitem [{\citenamefont {Urvoy}\ \emph {et~al.}(2015)\citenamefont {Urvoy},
  \citenamefont {Ripka}, \citenamefont {Lesanovsky}, \citenamefont {Booth},
  \citenamefont {Shaffer}, \citenamefont {Pfau},\ and\ \citenamefont
  {L{\"o}w}}]{urvoy2015}%
  \BibitemOpen
  \bibfield  {author} {\bibinfo {author} {\bibfnamefont {A.}~\bibnamefont
  {Urvoy}}, \bibinfo {author} {\bibfnamefont {F.}~\bibnamefont {Ripka}},
  \bibinfo {author} {\bibfnamefont {I.}~\bibnamefont {Lesanovsky}}, \bibinfo
  {author} {\bibfnamefont {D.}~\bibnamefont {Booth}}, \bibinfo {author}
  {\bibfnamefont {J.P.}\ \bibnamefont {Shaffer}}, \bibinfo {author}
  {\bibfnamefont {T.}~\bibnamefont {Pfau}}, \ and\ \bibinfo {author}
  {\bibfnamefont {R.}~\bibnamefont {L{\"o}w}},\ }\bibfield  {title} {\enquote
  {\bibinfo {title} {Strongly correlated growth of rydberg aggregates in a
  vapor cell},}\ }\href@noop {} {\bibfield  {journal} {\bibinfo  {journal}
  {Phys. Rev. Lett.}\ }\textbf {\bibinfo {volume} {114}},\ \bibinfo {pages}
  {203002} (\bibinfo {year} {2015})}\BibitemShut {NoStop}%
\bibitem [{\citenamefont {Schempp}\ \emph {et~al.}(2014)\citenamefont
  {Schempp}, \citenamefont {G{\"u}nter}, \citenamefont {Robert-de
  Saint-Vincent}, \citenamefont {Hofmann}, \citenamefont {Breyel},
  \citenamefont {Komnik}, \citenamefont {Sch{\"o}nleber}, \citenamefont
  {G{\"a}rttner}, \citenamefont {Evers}, \citenamefont {Whitlock},\ and\
  \citenamefont {Weidem{\"u}ller}}]{schempp2014}%
  \BibitemOpen
  \bibfield  {author} {\bibinfo {author} {\bibfnamefont {H.}~\bibnamefont
  {Schempp}}, \bibinfo {author} {\bibfnamefont {G.}~\bibnamefont {G{\"u}nter}},
  \bibinfo {author} {\bibfnamefont {M.}~\bibnamefont {Robert-de
  Saint-Vincent}}, \bibinfo {author} {\bibfnamefont {C.S.}\ \bibnamefont
  {Hofmann}}, \bibinfo {author} {\bibfnamefont {D.}~\bibnamefont {Breyel}},
  \bibinfo {author} {\bibfnamefont {A.}~\bibnamefont {Komnik}}, \bibinfo
  {author} {\bibfnamefont {D.W.}\ \bibnamefont {Sch{\"o}nleber}}, \bibinfo
  {author} {\bibfnamefont {M.}~\bibnamefont {G{\"a}rttner}}, \bibinfo {author}
  {\bibfnamefont {J.}~\bibnamefont {Evers}}, \bibinfo {author} {\bibfnamefont
  {S.}~\bibnamefont {Whitlock}}, \ and\ \bibinfo {author} {\bibfnamefont
  {M.}~\bibnamefont {Weidem{\"u}ller}},\ }\bibfield  {title} {\enquote
  {\bibinfo {title} {Full counting statistics of laser excited rydberg
  aggregates in a one-dimensional geometry},}\ }\href@noop {} {\bibfield
  {journal} {\bibinfo  {journal} {Phys. Rev. Lett.}\ }\textbf {\bibinfo
  {volume} {112}},\ \bibinfo {pages} {013002} (\bibinfo {year}
  {2014})}\BibitemShut {NoStop}%
\bibitem [{\citenamefont {Ates}\ \emph {et~al.}(2007)\citenamefont {Ates},
  \citenamefont {Pohl}, \citenamefont {Pattard},\ and\ \citenamefont
  {Rost}}]{ates2007}%
  \BibitemOpen
  \bibfield  {author} {\bibinfo {author} {\bibfnamefont {C.}~\bibnamefont
  {Ates}}, \bibinfo {author} {\bibfnamefont {T.}~\bibnamefont {Pohl}}, \bibinfo
  {author} {\bibfnamefont {T.}~\bibnamefont {Pattard}}, \ and\ \bibinfo
  {author} {\bibfnamefont {J.~M.}\ \bibnamefont {Rost}},\ }\bibfield  {title}
  {\enquote {\bibinfo {title} {Antiblockade in {R}ydberg excitation of an
  ultracold lattice gas},}\ }\href@noop {} {\bibfield  {journal} {\bibinfo
  {journal} {Phys. Rev. Lett.}\ }\textbf {\bibinfo {volume} {98}},\ \bibinfo
  {pages} {023002} (\bibinfo {year} {2007})}\BibitemShut {NoStop}%
\bibitem [{\citenamefont {Lesanovsky}\ and\ \citenamefont
  {Garrahan}(2013)}]{lesanovsky2013}%
  \BibitemOpen
  \bibfield  {author} {\bibinfo {author} {\bibfnamefont {I.}~\bibnamefont
  {Lesanovsky}}\ and\ \bibinfo {author} {\bibfnamefont {J.~P.}\ \bibnamefont
  {Garrahan}},\ }\bibfield  {title} {\enquote {\bibinfo {title} {Kinetic
  constraints, hierarchical relaxation, and onset of glassiness in strongly
  interacting and dissipative {R}ydberg gases},}\ }\href@noop {} {\bibfield
  {journal} {\bibinfo  {journal} {Phys. Rev. Lett.}\ }\textbf {\bibinfo
  {volume} {111}},\ \bibinfo {pages} {215305} (\bibinfo {year}
  {2013})}\BibitemShut {NoStop}%
\bibitem [{\citenamefont {G{\"a}rttner}\ \emph {et~al.}(2013)\citenamefont
  {G{\"a}rttner}, \citenamefont {Heeg}, \citenamefont {Gasenzer},\ and\
  \citenamefont {Evers}}]{garttner2013}%
  \BibitemOpen
  \bibfield  {author} {\bibinfo {author} {\bibfnamefont {M.}~\bibnamefont
  {G{\"a}rttner}}, \bibinfo {author} {\bibfnamefont {K.~P.}\ \bibnamefont
  {Heeg}}, \bibinfo {author} {\bibfnamefont {T.}~\bibnamefont {Gasenzer}}, \
  and\ \bibinfo {author} {\bibfnamefont {J.}~\bibnamefont {Evers}},\ }\bibfield
   {title} {\enquote {\bibinfo {title} {Dynamic formation of {R}ydberg
  aggregates at off-resonant excitation},}\ }\href@noop {} {\bibfield
  {journal} {\bibinfo  {journal} {Phys. Rev. A}\ }\textbf {\bibinfo {volume}
  {88}},\ \bibinfo {pages} {043410} (\bibinfo {year} {2013})}\BibitemShut
  {NoStop}%
\bibitem [{\citenamefont {Marcuzzi}\ \emph {et~al.}(2014)\citenamefont
  {Marcuzzi}, \citenamefont {Schick}, \citenamefont {Olmos},\ and\
  \citenamefont {Lesanovsky}}]{marcuzzi2014}%
  \BibitemOpen
  \bibfield  {author} {\bibinfo {author} {\bibfnamefont {M.}~\bibnamefont
  {Marcuzzi}}, \bibinfo {author} {\bibfnamefont {J.}~\bibnamefont {Schick}},
  \bibinfo {author} {\bibfnamefont {B.}~\bibnamefont {Olmos}}, \ and\ \bibinfo
  {author} {\bibfnamefont {I.}~\bibnamefont {Lesanovsky}},\ }\bibfield  {title}
  {\enquote {\bibinfo {title} {Effective dynamics of strongly dissipative
  {R}ydberg gases},}\ }\href@noop {} {\bibfield  {journal} {\bibinfo  {journal}
  {J. Phys. A: Math. Theor.}\ }\textbf {\bibinfo {volume} {47}},\ \bibinfo
  {pages} {482001} (\bibinfo {year} {2014})}\BibitemShut {NoStop}%
\bibitem [{\citenamefont {{\v{S}}ibali{\'c}}\ \emph {et~al.}(2016)\citenamefont
  {{\v{S}}ibali{\'c}}, \citenamefont {Wade}, \citenamefont {Adams},
  \citenamefont {Weatherill},\ and\ \citenamefont {Pohl}}]{sibalic2016}%
  \BibitemOpen
  \bibfield  {author} {\bibinfo {author} {\bibfnamefont {N.}~\bibnamefont
  {{\v{S}}ibali{\'c}}}, \bibinfo {author} {\bibfnamefont {C.~G.}\ \bibnamefont
  {Wade}}, \bibinfo {author} {\bibfnamefont {C.~S.}\ \bibnamefont {Adams}},
  \bibinfo {author} {\bibfnamefont {K.~J.}\ \bibnamefont {Weatherill}}, \ and\
  \bibinfo {author} {\bibfnamefont {T.}~\bibnamefont {Pohl}},\ }\bibfield
  {title} {\enquote {\bibinfo {title} {Driven-dissipative many-body systems
  with mixed power-law interactions: {B}istabilities and temperature-driven
  nonequilibrium phase transitions},}\ }\href@noop {} {\bibfield  {journal}
  {\bibinfo  {journal} {Phys. Rev. A}\ }\textbf {\bibinfo {volume} {94}},\
  \bibinfo {pages} {011401} (\bibinfo {year} {2016})}\BibitemShut {NoStop}%
\bibitem [{\citenamefont {Sekimoto}(1984)}]{sekimoto1984}%
  \BibitemOpen
  \bibfield  {author} {\bibinfo {author} {\bibfnamefont {K.}~\bibnamefont
  {Sekimoto}},\ }\bibfield  {title} {\enquote {\bibinfo {title} {Kinetics of
  magnetization switching in a 1-{D} system-size distribution of unswitched
  domains},}\ }\href@noop {} {\bibfield  {journal} {\bibinfo  {journal}
  {Physica A: Stat. Mech. Appl.}\ }\textbf {\bibinfo {volume} {125}},\ \bibinfo
  {pages} {261} (\bibinfo {year} {1984})}\BibitemShut {NoStop}%
\bibitem [{\citenamefont {Sekimoto}(1986)}]{sekimoto1986}%
  \BibitemOpen
  \bibfield  {author} {\bibinfo {author} {\bibfnamefont {K.}~\bibnamefont
  {Sekimoto}},\ }\bibfield  {title} {\enquote {\bibinfo {title} {Evolution of
  the domain structure during the nucleation-and-growth process with
  non-conserved order parameter},}\ }\href@noop {} {\bibfield  {journal}
  {\bibinfo  {journal} {Physica A: Stat. Mech. Appl.}\ }\textbf {\bibinfo
  {volume} {135}},\ \bibinfo {pages} {328} (\bibinfo {year}
  {1986})}\BibitemShut {NoStop}%
\bibitem [{\citenamefont {Sekimoto}(1991)}]{sekimoto1991}%
  \BibitemOpen
  \bibfield  {author} {\bibinfo {author} {\bibfnamefont {K.}~\bibnamefont
  {Sekimoto}},\ }\bibfield  {title} {\enquote {\bibinfo {title} {Evolution of
  the domain structure during the nucleation-and-growth process with
  non-conserved order parameter},}\ }\href@noop {} {\bibfield  {journal}
  {\bibinfo  {journal} {Int. J. Mod. Phys. B}\ }\textbf {\bibinfo {volume}
  {5}},\ \bibinfo {pages} {1843} (\bibinfo {year} {1991})}\BibitemShut
  {NoStop}%
\bibitem [{\citenamefont {Ben-Naim}\ and\ \citenamefont
  {Krapivsky}(1996)}]{benami1996}%
  \BibitemOpen
  \bibfield  {author} {\bibinfo {author} {\bibfnamefont {E.}~\bibnamefont
  {Ben-Naim}}\ and\ \bibinfo {author} {\bibfnamefont {P.~L.}\ \bibnamefont
  {Krapivsky}},\ }\bibfield  {title} {\enquote {\bibinfo {title} {Nucleation
  and growth in one dimension},}\ }\href@noop {} {\bibfield  {journal}
  {\bibinfo  {journal} {Phys. Rev. E}\ }\textbf {\bibinfo {volume} {54}},\
  \bibinfo {pages} {3562} (\bibinfo {year} {1996})}\BibitemShut {NoStop}%
\bibitem [{\citenamefont {Bloch}\ \emph {et~al.}(2012)\citenamefont {Bloch},
  \citenamefont {Dalibard},\ and\ \citenamefont {Nascimbene}}]{bloch2012}%
  \BibitemOpen
  \bibfield  {author} {\bibinfo {author} {\bibfnamefont {I.}~\bibnamefont
  {Bloch}}, \bibinfo {author} {\bibfnamefont {J.}~\bibnamefont {Dalibard}}, \
  and\ \bibinfo {author} {\bibfnamefont {S.}~\bibnamefont {Nascimbene}},\
  }\bibfield  {title} {\enquote {\bibinfo {title} {Quantum simulations with
  ultracold quantum gases},}\ }\href@noop {} {\bibfield  {journal} {\bibinfo
  {journal} {Nat. Phys.}\ }\textbf {\bibinfo {volume} {8}},\ \bibinfo {pages}
  {267} (\bibinfo {year} {2012})}\BibitemShut {NoStop}%
\bibitem [{\citenamefont {Labuhn}\ \emph {et~al.}(2016)\citenamefont {Labuhn},
  \citenamefont {Barredo}, \citenamefont {Ravets}, \citenamefont
  {De~L{\'e}s{\'e}leuc}, \citenamefont {Macr{\`\i}}, \citenamefont {Lahaye},\
  and\ \citenamefont {Browaeys}}]{labuhn2016}%
  \BibitemOpen
  \bibfield  {author} {\bibinfo {author} {\bibfnamefont {H.}~\bibnamefont
  {Labuhn}}, \bibinfo {author} {\bibfnamefont {D.}~\bibnamefont {Barredo}},
  \bibinfo {author} {\bibfnamefont {S.}~\bibnamefont {Ravets}}, \bibinfo
  {author} {\bibfnamefont {S.}~\bibnamefont {De~L{\'e}s{\'e}leuc}}, \bibinfo
  {author} {\bibfnamefont {T.}~\bibnamefont {Macr{\`\i}}}, \bibinfo {author}
  {\bibfnamefont {T.}~\bibnamefont {Lahaye}}, \ and\ \bibinfo {author}
  {\bibfnamefont {A.}~\bibnamefont {Browaeys}},\ }\bibfield  {title} {\enquote
  {\bibinfo {title} {Tunable two-dimensional arrays of single {R}ydberg atoms
  for realizing quantum {I}sing models},}\ }\href@noop {} {\bibfield  {journal}
  {\bibinfo  {journal} {Nature}\ }\textbf {\bibinfo {volume} {534}},\ \bibinfo
  {pages} {667} (\bibinfo {year} {2016})}\BibitemShut {NoStop}%
\bibitem [{\citenamefont {Bernien}\ \emph {et~al.}(2017)\citenamefont
  {Bernien}, \citenamefont {Schwartz}, \citenamefont {Keesling}, \citenamefont
  {Levine}, \citenamefont {Omran}, \citenamefont {Pichler}, \citenamefont
  {Choi}, \citenamefont {Zibrov}, \citenamefont {Endres}, \citenamefont
  {Greiner}, \citenamefont {Vuletić},\ and\ \citenamefont
  {Lukin}}]{bernien2017}%
  \BibitemOpen
  \bibfield  {author} {\bibinfo {author} {\bibfnamefont {H.}~\bibnamefont
  {Bernien}}, \bibinfo {author} {\bibfnamefont {S.}~\bibnamefont {Schwartz}},
  \bibinfo {author} {\bibfnamefont {A.}~\bibnamefont {Keesling}}, \bibinfo
  {author} {\bibfnamefont {H.}~\bibnamefont {Levine}}, \bibinfo {author}
  {\bibfnamefont {A.}~\bibnamefont {Omran}}, \bibinfo {author} {\bibfnamefont
  {H.}~\bibnamefont {Pichler}}, \bibinfo {author} {\bibfnamefont
  {S.}~\bibnamefont {Choi}}, \bibinfo {author} {\bibfnamefont {A.~S.}\
  \bibnamefont {Zibrov}}, \bibinfo {author} {\bibfnamefont {M.}~\bibnamefont
  {Endres}}, \bibinfo {author} {\bibfnamefont {M.}~\bibnamefont {Greiner}},
  \bibinfo {author} {\bibfnamefont {V.}~\bibnamefont {Vuletić}}, \ and\
  \bibinfo {author} {\bibfnamefont {M.~D.}\ \bibnamefont {Lukin}},\ }\bibfield
  {title} {\enquote {\bibinfo {title} {Probing many-body dynamics on a 51-atom
  quantum simulator},}\ }\href@noop {} {\bibfield  {journal} {\bibinfo
  {journal} {Nature}\ }\textbf {\bibinfo {volume} {551}},\ \bibinfo {pages}
  {579} (\bibinfo {year} {2017})}\BibitemShut {NoStop}%
\bibitem [{\citenamefont {Lienhard}\ \emph {et~al.}(2017)\citenamefont
  {Lienhard}, \citenamefont {de~L{\'e}s{\'e}leuc}, \citenamefont {Barredo},
  \citenamefont {Lahaye}, \citenamefont {Browaeys}, \citenamefont {Schuler},
  \citenamefont {Henry},\ and\ \citenamefont {L{\"a}uchli}}]{lienhard2017}%
  \BibitemOpen
  \bibfield  {author} {\bibinfo {author} {\bibfnamefont {V.}~\bibnamefont
  {Lienhard}}, \bibinfo {author} {\bibfnamefont {S.}~\bibnamefont
  {de~L{\'e}s{\'e}leuc}}, \bibinfo {author} {\bibfnamefont {D.}~\bibnamefont
  {Barredo}}, \bibinfo {author} {\bibfnamefont {T.}~\bibnamefont {Lahaye}},
  \bibinfo {author} {\bibfnamefont {A.}~\bibnamefont {Browaeys}}, \bibinfo
  {author} {\bibfnamefont {M.}~\bibnamefont {Schuler}}, \bibinfo {author}
  {\bibfnamefont {L.-P.}\ \bibnamefont {Henry}}, \ and\ \bibinfo {author}
  {\bibfnamefont {A.~M.}\ \bibnamefont {L{\"a}uchli}},\ }\bibfield  {title}
  {\enquote {\bibinfo {title} {Observing the space-and time-dependent growth of
  correlations in dynamically tuned synthetic {I}sing antiferromagnets},}\
  }\href@noop {} {\bibfield  {journal} {\bibinfo  {journal} {arXiv:1711.01185}\
  } (\bibinfo {year} {2017})}\BibitemShut {NoStop}%
\bibitem [{\citenamefont {Zeiher}\ \emph {et~al.}(2017)\citenamefont {Zeiher},
  \citenamefont {Choi}, \citenamefont {Rubio-Abadal}, \citenamefont {Pohl},
  \citenamefont {van Bijnen}, \citenamefont {Bloch},\ and\ \citenamefont
  {Gross}}]{zeiher2017}%
  \BibitemOpen
  \bibfield  {author} {\bibinfo {author} {\bibfnamefont {J.}~\bibnamefont
  {Zeiher}}, \bibinfo {author} {\bibfnamefont {J.}~\bibnamefont {Choi}},
  \bibinfo {author} {\bibfnamefont {A.}~\bibnamefont {Rubio-Abadal}}, \bibinfo
  {author} {\bibfnamefont {T.}~\bibnamefont {Pohl}}, \bibinfo {author}
  {\bibfnamefont {R.}~\bibnamefont {van Bijnen}}, \bibinfo {author}
  {\bibfnamefont {I.}~\bibnamefont {Bloch}}, \ and\ \bibinfo {author}
  {\bibfnamefont {C.}~\bibnamefont {Gross}},\ }\bibfield  {title} {\enquote
  {\bibinfo {title} {Coherent many-body spin dynamics in a long-range
  interacting {I}sing chain},}\ }\href@noop {} {\bibfield  {journal} {\bibinfo
  {journal} {Phys. Rev. X}\ }\textbf {\bibinfo {volume} {7}},\ \bibinfo {pages}
  {041063} (\bibinfo {year} {2017})}\BibitemShut {NoStop}%
\bibitem [{\citenamefont {Helmrich}\ \emph {et~al.}(2016)\citenamefont
  {Helmrich}, \citenamefont {Arias},\ and\ \citenamefont
  {Whitlock}}]{helmrich2016}%
  \BibitemOpen
  \bibfield  {author} {\bibinfo {author} {\bibfnamefont {S.}~\bibnamefont
  {Helmrich}}, \bibinfo {author} {\bibfnamefont {A.}~\bibnamefont {Arias}}, \
  and\ \bibinfo {author} {\bibfnamefont {S.}~\bibnamefont {Whitlock}},\
  }\bibfield  {title} {\enquote {\bibinfo {title} {Scaling of a long-range
  interacting quantum spin system driven out of equilibrium},}\ }\href@noop {}
  {\bibfield  {journal} {\bibinfo  {journal} {arXiv:1605.08609}\ } (\bibinfo
  {year} {2016})}\BibitemShut {NoStop}%
\bibitem [{\citenamefont {Letscher}\ \emph {et~al.}(2017)\citenamefont
  {Letscher}, \citenamefont {Thomas}, \citenamefont {Niederpr{\"u}m},
  \citenamefont {Fleischhauer},\ and\ \citenamefont {Ott}}]{letscher2017}%
  \BibitemOpen
  \bibfield  {author} {\bibinfo {author} {\bibfnamefont {F.}~\bibnamefont
  {Letscher}}, \bibinfo {author} {\bibfnamefont {O.}~\bibnamefont {Thomas}},
  \bibinfo {author} {\bibfnamefont {T.}~\bibnamefont {Niederpr{\"u}m}},
  \bibinfo {author} {\bibfnamefont {M.}~\bibnamefont {Fleischhauer}}, \ and\
  \bibinfo {author} {\bibfnamefont {H.}~\bibnamefont {Ott}},\ }\bibfield
  {title} {\enquote {\bibinfo {title} {Bistability versus metastability in
  driven dissipative {R}ydberg gases},}\ }\href@noop {} {\bibfield  {journal}
  {\bibinfo  {journal} {Phys. Rev. X}\ }\textbf {\bibinfo {volume} {7}},\
  \bibinfo {pages} {021020} (\bibinfo {year} {2017})}\BibitemShut {NoStop}%
\bibitem [{\citenamefont {Valado}\ \emph {et~al.}(2016)\citenamefont {Valado},
  \citenamefont {Simonelli}, \citenamefont {Hoogerland}, \citenamefont
  {Lesanovsky}, \citenamefont {Garrahan}, \citenamefont {Arimondo},
  \citenamefont {Ciampini},\ and\ \citenamefont {Morsch}}]{valado2016}%
  \BibitemOpen
  \bibfield  {author} {\bibinfo {author} {\bibfnamefont {M.~M.}\ \bibnamefont
  {Valado}}, \bibinfo {author} {\bibfnamefont {C.}~\bibnamefont {Simonelli}},
  \bibinfo {author} {\bibfnamefont {M.~D.}\ \bibnamefont {Hoogerland}},
  \bibinfo {author} {\bibfnamefont {I.}~\bibnamefont {Lesanovsky}}, \bibinfo
  {author} {\bibfnamefont {J.~P.}\ \bibnamefont {Garrahan}}, \bibinfo {author}
  {\bibfnamefont {E.}~\bibnamefont {Arimondo}}, \bibinfo {author}
  {\bibfnamefont {D.}~\bibnamefont {Ciampini}}, \ and\ \bibinfo {author}
  {\bibfnamefont {O.}~\bibnamefont {Morsch}},\ }\bibfield  {title} {\enquote
  {\bibinfo {title} {Experimental observation of controllable kinetic
  constraints in a cold atomic gas},}\ }\href@noop {} {\bibfield  {journal}
  {\bibinfo  {journal} {Phys. Rev. A}\ }\textbf {\bibinfo {volume} {93}},\
  \bibinfo {pages} {040701} (\bibinfo {year} {2016})}\BibitemShut {NoStop}%
\bibitem [{SM()}]{SM}%
  \BibitemOpen
  \href@noop {} {}\bibinfo {note} {See the Supplemental Material for
  details.}\BibitemShut {Stop}%
\bibitem [{\citenamefont {Guti{\'e}rrez}\ \emph {et~al.}(2015)\citenamefont
  {Guti{\'e}rrez}, \citenamefont {Garrahan},\ and\ \citenamefont
  {Lesanovsky}}]{gutierrez2015}%
  \BibitemOpen
  \bibfield  {author} {\bibinfo {author} {\bibfnamefont {R.}~\bibnamefont
  {Guti{\'e}rrez}}, \bibinfo {author} {\bibfnamefont {J.~P.}\ \bibnamefont
  {Garrahan}}, \ and\ \bibinfo {author} {\bibfnamefont {I.}~\bibnamefont
  {Lesanovsky}},\ }\bibfield  {title} {\enquote {\bibinfo {title} {Self-similar
  nonequilibrium dynamics of a many-body system with power-law interactions},}\
  }\href@noop {} {\bibfield  {journal} {\bibinfo  {journal} {Phys. Rev. E}\
  }\textbf {\bibinfo {volume} {92}},\ \bibinfo {pages} {062144} (\bibinfo
  {year} {2015})}\BibitemShut {NoStop}%
\bibitem [{\citenamefont {Bortz}\ \emph {et~al.}(1975)\citenamefont {Bortz},
  \citenamefont {Kalos},\ and\ \citenamefont {Lebowitz}}]{bortz1975}%
  \BibitemOpen
  \bibfield  {author} {\bibinfo {author} {\bibfnamefont {A.~B.}\ \bibnamefont
  {Bortz}}, \bibinfo {author} {\bibfnamefont {M.~H.}\ \bibnamefont {Kalos}}, \
  and\ \bibinfo {author} {\bibfnamefont {J.~L.}\ \bibnamefont {Lebowitz}},\
  }\bibfield  {title} {\enquote {\bibinfo {title} {A new algorithm for monte
  carlo simulation of ising spin systems},}\ }\href@noop {} {\bibfield
  {journal} {\bibinfo  {journal} {J. Comput. Phys.}\ }\textbf {\bibinfo
  {volume} {17}},\ \bibinfo {pages} {10} (\bibinfo {year} {1975})}\BibitemShut
  {NoStop}%
\bibitem [{\citenamefont {Marcuzzi}\ \emph {et~al.}(2015)\citenamefont
  {Marcuzzi}, \citenamefont {Levi}, \citenamefont {Li}, \citenamefont
  {Garrahan}, \citenamefont {Olmos},\ and\ \citenamefont
  {Lesanovsky}}]{marcuzzi2015}%
  \BibitemOpen
  \bibfield  {author} {\bibinfo {author} {\bibfnamefont {M.}~\bibnamefont
  {Marcuzzi}}, \bibinfo {author} {\bibfnamefont {E.}~\bibnamefont {Levi}},
  \bibinfo {author} {\bibfnamefont {W.}~\bibnamefont {Li}}, \bibinfo {author}
  {\bibfnamefont {J.~P.}\ \bibnamefont {Garrahan}}, \bibinfo {author}
  {\bibfnamefont {B.}~\bibnamefont {Olmos}}, \ and\ \bibinfo {author}
  {\bibfnamefont {I.}~\bibnamefont {Lesanovsky}},\ }\bibfield  {title}
  {\enquote {\bibinfo {title} {Non-equilibrium universality in the dynamics of
  dissipative cold atomic gases},}\ }\href@noop {} {\bibfield  {journal}
  {\bibinfo  {journal} {New J. Phys.}\ }\textbf {\bibinfo {volume} {17}},\
  \bibinfo {pages} {072003} (\bibinfo {year} {2015})}\BibitemShut {NoStop}%
\bibitem [{\citenamefont {Guti{\'e}rrez}\ \emph {et~al.}(2017)\citenamefont
  {Guti{\'e}rrez}, \citenamefont {Simonelli}, \citenamefont {Archimi},
  \citenamefont {Castellucci}, \citenamefont {Arimondo}, \citenamefont
  {Ciampini}, \citenamefont {Marcuzzi}, \citenamefont {Lesanovsky},\ and\
  \citenamefont {Morsch}}]{gutierrez2016c}%
  \BibitemOpen
  \bibfield  {author} {\bibinfo {author} {\bibfnamefont {R.}~\bibnamefont
  {Guti{\'e}rrez}}, \bibinfo {author} {\bibfnamefont {C.}~\bibnamefont
  {Simonelli}}, \bibinfo {author} {\bibfnamefont {M.}~\bibnamefont {Archimi}},
  \bibinfo {author} {\bibfnamefont {F.}~\bibnamefont {Castellucci}}, \bibinfo
  {author} {\bibfnamefont {E.}~\bibnamefont {Arimondo}}, \bibinfo {author}
  {\bibfnamefont {D.}~\bibnamefont {Ciampini}}, \bibinfo {author}
  {\bibfnamefont {M.}~\bibnamefont {Marcuzzi}}, \bibinfo {author}
  {\bibfnamefont {I.}~\bibnamefont {Lesanovsky}}, \ and\ \bibinfo {author}
  {\bibfnamefont {O.}~\bibnamefont {Morsch}},\ }\bibfield  {title} {\enquote
  {\bibinfo {title} {Experimental signatures of an absorbing-state phase
  transition in an open driven many-body quantum system},}\ }\href@noop {}
  {\bibfield  {journal} {\bibinfo  {journal} {Physical Review A}\ }\textbf
  {\bibinfo {volume} {96}},\ \bibinfo {pages} {041602} (\bibinfo {year}
  {2017})}\BibitemShut {NoStop}%
\bibitem [{\citenamefont {Marcuzzi}\ \emph {et~al.}(2016)\citenamefont
  {Marcuzzi}, \citenamefont {Buchhold}, \citenamefont {Diehl},\ and\
  \citenamefont {Lesanovsky}}]{marcuzzi2016}%
  \BibitemOpen
  \bibfield  {author} {\bibinfo {author} {\bibfnamefont {M.}~\bibnamefont
  {Marcuzzi}}, \bibinfo {author} {\bibfnamefont {M.}~\bibnamefont {Buchhold}},
  \bibinfo {author} {\bibfnamefont {S.}~\bibnamefont {Diehl}}, \ and\ \bibinfo
  {author} {\bibfnamefont {I.}~\bibnamefont {Lesanovsky}},\ }\bibfield  {title}
  {\enquote {\bibinfo {title} {Absorbing state phase transition with competing
  quantum and classical fluctuations},}\ }\href@noop {} {\bibfield  {journal}
  {\bibinfo  {journal} {Phys. Rev. Lett.}\ }\textbf {\bibinfo {volume} {116}},\
  \bibinfo {pages} {245701} (\bibinfo {year} {2016})}\BibitemShut {NoStop}%
\bibitem [{\citenamefont {Dalibard}\ \emph {et~al.}(1992)\citenamefont
  {Dalibard}, \citenamefont {Castin},\ and\ \citenamefont
  {M{\o}lmer}}]{dalibard1992}%
  \BibitemOpen
  \bibfield  {author} {\bibinfo {author} {\bibfnamefont {J.}~\bibnamefont
  {Dalibard}}, \bibinfo {author} {\bibfnamefont {Y.}~\bibnamefont {Castin}}, \
  and\ \bibinfo {author} {\bibfnamefont {K.}~\bibnamefont {M{\o}lmer}},\
  }\bibfield  {title} {\enquote {\bibinfo {title} {Wave-function approach to
  dissipative processes in quantum optics},}\ }\href@noop {} {\bibfield
  {journal} {\bibinfo  {journal} {Phys. Rev. Lett.}\ }\textbf {\bibinfo
  {volume} {68}},\ \bibinfo {pages} {580} (\bibinfo {year} {1992})}\BibitemShut
  {NoStop}%
\bibitem [{\citenamefont {Everest}\ \emph {et~al.}(2017)\citenamefont
  {Everest}, \citenamefont {Lesanovsky}, \citenamefont {Garrahan},\ and\
  \citenamefont {Levi}}]{everest2017}%
  \BibitemOpen
  \bibfield  {author} {\bibinfo {author} {\bibfnamefont {B.}~\bibnamefont
  {Everest}}, \bibinfo {author} {\bibfnamefont {I.}~\bibnamefont {Lesanovsky}},
  \bibinfo {author} {\bibfnamefont {J.~P.}\ \bibnamefont {Garrahan}}, \ and\
  \bibinfo {author} {\bibfnamefont {E.}~\bibnamefont {Levi}},\ }\bibfield
  {title} {\enquote {\bibinfo {title} {Role of interactions in a dissipative
  many-body localized system},}\ }\href@noop {} {\bibfield  {journal} {\bibinfo
   {journal} {Phys. Rev. B}\ }\textbf {\bibinfo {volume} {95}},\ \bibinfo
  {pages} {024310} (\bibinfo {year} {2017})}\BibitemShut {NoStop}%
\end{thebibliography}

\begin{thebibliography}{12}%
\makeatletter
\providecommand \@ifxundefined [1]{%
 \@ifx{#1\undefined}
}%
\providecommand \@ifnum [1]{%
 \ifnum #1\expandafter \@firstoftwo
 \else \expandafter \@secondoftwo
 \fi
}%
\providecommand \@ifx [1]{%
 \ifx #1\expandafter \@firstoftwo
 \else \expandafter \@secondoftwo
 \fi
}%
\providecommand \natexlab [1]{#1}%
\providecommand \enquote  [1]{``#1''}%
\providecommand \bibnamefont  [1]{#1}%
\providecommand \bibfnamefont [1]{#1}%
\providecommand \citenamefont [1]{#1}%
\providecommand \href@noop [0]{\@secondoftwo}%
\providecommand \href [0]{\begingroup \@sanitize@url \@href}%
\providecommand \@href[1]{\@@startlink{#1}\@@href}%
\providecommand \@@href[1]{\endgroup#1\@@endlink}%
\providecommand \@sanitize@url [0]{\catcode `\\12\catcode `\$12\catcode
  `\&12\catcode `\#12\catcode `\^12\catcode `\_12\catcode `\%12\relax}%
\providecommand \@@startlink[1]{}%
\providecommand \@@endlink[0]{}%
\providecommand \url  [0]{\begingroup\@sanitize@url \@url }%
\providecommand \@url [1]{\endgroup\@href {#1}{\urlprefix }}%
\providecommand \urlprefix  [0]{URL }%
\providecommand \Eprint [0]{\href }%
\providecommand \doibase [0]{http://dx.doi.org/}%
\providecommand \selectlanguage [0]{\@gobble}%
\providecommand \bibinfo  [0]{\@secondoftwo}%
\providecommand \bibfield  [0]{\@secondoftwo}%
\providecommand \translation [1]{[#1]}%
\providecommand \BibitemOpen [0]{}%
\providecommand \bibitemStop [0]{}%
\providecommand \bibitemNoStop [0]{.\EOS\space}%
\providecommand \EOS [0]{\spacefactor3000\relax}%
\providecommand \BibitemShut  [1]{\csname bibitem#1\endcsname}%
\let\auto@bib@innerbib\@empty
\bibitem [{\citenamefont {Kolmogorov}(1937)}]{kolmogorov1937SM}%
  \BibitemOpen
  \bibfield  {author} {\bibinfo {author} {\bibfnamefont {A.~E.}\ \bibnamefont
  {Kolmogorov}},\ }\bibfield  {title} {\enquote {\bibinfo {title} {On the
  statistical theory of metal crystallization},}\ }\href@noop {} {\bibfield
  {journal} {\bibinfo  {journal} {Izv. Akad. Nauk. SSSR Ser. Mat.}\ }\textbf
  {\bibinfo {volume} {1}},\ \bibinfo {pages} {333} (\bibinfo {year}
  {1937})}\BibitemShut {NoStop}%
\bibitem [{\citenamefont {Avrami}(1939)}]{avrami1939SM}%
  \BibitemOpen
  \bibfield  {author} {\bibinfo {author} {\bibfnamefont {M.}~\bibnamefont
  {Avrami}},\ }\bibfield  {title} {\enquote {\bibinfo {title} {Kinetics of
  phase change. {I} {G}eneral theory},}\ }\href@noop {} {\bibfield  {journal}
  {\bibinfo  {journal} {J. Chem. Phys.}\ }\textbf {\bibinfo {volume} {7}},\
  \bibinfo {pages} {1103} (\bibinfo {year} {1939})}\BibitemShut {NoStop}%
\bibitem [{\citenamefont {Johnson}\ and\ \citenamefont
  {Mehl}(1939)}]{johnson1939SM}%
  \BibitemOpen
  \bibfield  {author} {\bibinfo {author} {\bibfnamefont {W.~A.}\ \bibnamefont
  {Johnson}}\ and\ \bibinfo {author} {\bibfnamefont {R.~F.}\ \bibnamefont
  {Mehl}},\ }\bibfield  {title} {\enquote {\bibinfo {title} {Reaction kinetics
  in processes of nucleation and growth},}\ }\href@noop {} {\bibfield
  {journal} {\bibinfo  {journal} {Trans. Am. Inst. Min. Eng.}\ }\textbf
  {\bibinfo {volume} {135}},\ \bibinfo {pages} {416} (\bibinfo {year}
  {1939})}\BibitemShut {NoStop}%
\bibitem [{\citenamefont {Avrami}(1940)}]{avrami1940SM}%
  \BibitemOpen
  \bibfield  {author} {\bibinfo {author} {\bibfnamefont {M.}~\bibnamefont
  {Avrami}},\ }\bibfield  {title} {\enquote {\bibinfo {title} {Kinetics of
  phase change. {II} {T}ransformation-time relations for random distribution of
  nuclei},}\ }\href@noop {} {\bibfield  {journal} {\bibinfo  {journal} {J.
  Chem. Phys.}\ }\textbf {\bibinfo {volume} {8}},\ \bibinfo {pages} {212}
  (\bibinfo {year} {1940})}\BibitemShut {NoStop}%
\bibitem [{\citenamefont {Avrami}(1941)}]{avrami1941SM}%
  \BibitemOpen
  \bibfield  {author} {\bibinfo {author} {\bibfnamefont {M.}~\bibnamefont
  {Avrami}},\ }\bibfield  {title} {\enquote {\bibinfo {title} {Granulation,
  phase change, and microstructure kinetics of phase change. {III}},}\
  }\href@noop {} {\bibfield  {journal} {\bibinfo  {journal} {J. Chem. Phys.}\
  }\textbf {\bibinfo {volume} {9}},\ \bibinfo {pages} {177} (\bibinfo {year}
  {1941})}\BibitemShut {NoStop}%
\bibitem [{\citenamefont {Sekimoto}(1984)}]{sekimoto1984SM}%
  \BibitemOpen
  \bibfield  {author} {\bibinfo {author} {\bibfnamefont {K.}~\bibnamefont
  {Sekimoto}},\ }\bibfield  {title} {\enquote {\bibinfo {title} {Kinetics of
  magnetization switching in a 1-{D} system-size distribution of unswitched
  domains},}\ }\href@noop {} {\bibfield  {journal} {\bibinfo  {journal}
  {Physica A: Stat. Mech. Appl.}\ }\textbf {\bibinfo {volume} {125}},\ \bibinfo
  {pages} {261} (\bibinfo {year} {1984})}\BibitemShut {NoStop}%
\bibitem [{\citenamefont {Sekimoto}(1986)}]{sekimoto1986SM}%
  \BibitemOpen
  \bibfield  {author} {\bibinfo {author} {\bibfnamefont {K.}~\bibnamefont
  {Sekimoto}},\ }\bibfield  {title} {\enquote {\bibinfo {title} {Evolution of
  the domain structure during the nucleation-and-growth process with
  non-conserved order parameter},}\ }\href@noop {} {\bibfield  {journal}
  {\bibinfo  {journal} {Physica A: Stat. Mech. Appl.}\ }\textbf {\bibinfo
  {volume} {135}},\ \bibinfo {pages} {328} (\bibinfo {year}
  {1986})}\BibitemShut {NoStop}%
\bibitem [{\citenamefont {Sekimoto}(1991)}]{sekimoto1991SM}%
  \BibitemOpen
  \bibfield  {author} {\bibinfo {author} {\bibfnamefont {K.}~\bibnamefont
  {Sekimoto}},\ }\bibfield  {title} {\enquote {\bibinfo {title} {Evolution of
  the domain structure during the nucleation-and-growth process with
  non-conserved order parameter},}\ }\href@noop {} {\bibfield  {journal}
  {\bibinfo  {journal} {Int. J. Mod. Phys. B}\ }\textbf {\bibinfo {volume}
  {5}},\ \bibinfo {pages} {1843} (\bibinfo {year} {1991})}\BibitemShut
  {NoStop}%
\bibitem [{\citenamefont {Ben-Naim}\ and\ \citenamefont
  {Krapivsky}(1996)}]{benami1996SM}%
  \BibitemOpen
  \bibfield  {author} {\bibinfo {author} {\bibfnamefont {E.}~\bibnamefont
  {Ben-Naim}}\ and\ \bibinfo {author} {\bibfnamefont {P.~L.}\ \bibnamefont
  {Krapivsky}},\ }\bibfield  {title} {\enquote {\bibinfo {title} {Nucleation
  and growth in one dimension},}\ }\href@noop {} {\bibfield  {journal}
  {\bibinfo  {journal} {Phys. Rev. E}\ }\textbf {\bibinfo {volume} {54}},\
  \bibinfo {pages} {3562} (\bibinfo {year} {1996})}\BibitemShut {NoStop}%
\bibitem [{\citenamefont {Jun}\ \emph {et~al.}(2005)\citenamefont {Jun},
  \citenamefont {Zhang},\ and\ \citenamefont {Bechhoefer}}]{jun2005aSM}%
  \BibitemOpen
  \bibfield  {author} {\bibinfo {author} {\bibfnamefont {S.}~\bibnamefont
  {Jun}}, \bibinfo {author} {\bibfnamefont {H.}~\bibnamefont {Zhang}}, \ and\
  \bibinfo {author} {\bibfnamefont {J.}~\bibnamefont {Bechhoefer}},\ }\bibfield
   {title} {\enquote {\bibinfo {title} {Nucleation and growth in one dimension.
  {I}. {T}he generalized {K}olmogorov-{J}ohnson-{M}ehl-{A}vrami model},}\
  }\href@noop {} {\bibfield  {journal} {\bibinfo  {journal} {Phys. Rev. E}\
  }\textbf {\bibinfo {volume} {71}},\ \bibinfo {pages} {011908} (\bibinfo
  {year} {2005})}\BibitemShut {NoStop}%
\bibitem [{\citenamefont {Dalibard}\ \emph {et~al.}(1992)\citenamefont
  {Dalibard}, \citenamefont {Castin},\ and\ \citenamefont
  {M{\o}lmer}}]{dalibard1992SM}%
  \BibitemOpen
  \bibfield  {author} {\bibinfo {author} {\bibfnamefont {J.}~\bibnamefont
  {Dalibard}}, \bibinfo {author} {\bibfnamefont {Y.}~\bibnamefont {Castin}}, \
  and\ \bibinfo {author} {\bibfnamefont {K.}~\bibnamefont {M{\o}lmer}},\
  }\bibfield  {title} {\enquote {\bibinfo {title} {Wave-function approach to
  dissipative processes in quantum optics},}\ }\href@noop {} {\bibfield
  {journal} {\bibinfo  {journal} {Phys. Rev. Lett.}\ }\textbf {\bibinfo
  {volume} {68}},\ \bibinfo {pages} {580} (\bibinfo {year} {1992})}\BibitemShut
  {NoStop}%
\bibitem [{\citenamefont {Plenio}\ and\ \citenamefont
  {Knight}(1998)}]{plenio1998SM}%
  \BibitemOpen
  \bibfield  {author} {\bibinfo {author} {\bibfnamefont {M.B.}\ \bibnamefont
  {Plenio}}\ and\ \bibinfo {author} {\bibfnamefont {P.L.}\ \bibnamefont
  {Knight}},\ }\bibfield  {title} {\enquote {\bibinfo {title} {The quantum-jump
  approach to dissipative dynamics in quantum optics},}\ }\href@noop {}
  {\bibfield  {journal} {\bibinfo  {journal} {Rev. Mod. Phys.}\ }\textbf
  {\bibinfo {volume} {70}},\ \bibinfo {pages} {101} (\bibinfo {year}
  {1998})}\BibitemShut {NoStop}%
\end{thebibliography}
%

\end{document}